\documentclass[%
 reprint,
 amsmath,amssymb,
 aps,nofootinbib,
]{revtex4-2}

\usepackage{graphicx}% Include figure files
\usepackage{dcolumn}% Align table columns on decimal point
\usepackage{bm}% bold math
\usepackage{braket}
\usepackage{lipsum}
\usepackage[dvipsnames]{xcolor}

\begin{document}

\preprint{APS/123-QED}

\title{Error Correlations in Photonic Qudit-Mediated Entanglement Generation}% Force line breaks with \\

\author{Xiaoyu Liu$^{1,2}$}%
 \email{xiaoyu@lorentz.leidenuniv.nl}
\author{Niv Bharos$^{2, 3}$}%
 \email{niv@mit.edu}
\author{Liubov Markovich$^{1,2,4,5}$}%
 \email{markovich@lorentz.leidenuniv.nl}
\author{Johannes Borregaard$^{2,6}$}%
 \email{borregaard@fas.harvard.edu}
 \affiliation{$^1$Instituut-Lorentz, Universiteit Leiden, Niels Bohrweg 2, 2333CA Leiden, Netherlands}%
\affiliation{$^2$QuTech and Kavli Institute of Nanoscience, Technische Universiteit Delft, Lorentzweg 1, 2628CJ Delft, Netherlands}%
\affiliation{$^3$Research Laboratory of Electronics, Massachusetts Institute of Technology, Cambridge, Massachusetts 02139, USA}
\affiliation{$^4$Institute for Information Transmission Problems,  Bol. Karetny per. 19, Moscow 127051, Russia}
\affiliation{$^5$Russian Quantum Center, Skolkovo, Moscow 121205, Russia}
\affiliation{$^6$Department of Physics, Harvard University, 17 Oxford Street Cambridge, Massachusetts 021388, USA}

\date{\today}% It is always \today, today,

\begin{abstract}
Generating entanglement between distributed network nodes is a prerequisite for the quantum internet. Entanglement distribution protocols based on high-dimensional photonic qudits enable the simultaneous generation of multiple entangled pairs, which can significantly reduce the required coherence time of the qubit registers. However, current schemes require fast optical switching, which is experimentally challenging. In addition, the higher degree of error correlation between the generated entangled pairs in qudit protocols compared to qubit protocols has not been studied in detail. We propose a qudit-mediated entangling protocol that completely circumvents the need for optical switches, making it more accessible for current experimental systems. Furthermore, we quantify the amount of error correlation between the simultaneously generated entangled pairs and analyze the effect on entanglement purification algorithms and teleportation-based quantum error correction. We find that optimized purification schemes can efficiently correct the correlated errors, while the quantum error correction codes studied here perform worse than for uncorrelated error models.

\end{abstract}

\maketitle

%\tableofcontents

\section{\label{sec:level1}Introduction}

Quantum networking enables new primitives such as information-theoretically secure communication~\cite{pirandola2020advances}, quantum sensing networks~\cite{degen2017quantum, zhang2021distributed,guo2020distributed}, and distributed quantum computation~\cite{beals2013efficient,cacciapuoti2019quantum, buhrman2003distributed}. Furthermore, it provides a promising route towards scalable quantum computers via modular designs~\cite{Awschalom2021}. 
A prerequisite for these applications is the distribution of high-fidelity entanglement between the network nodes to enable reliable transfer of quantum information by quantum teleportation. 

Entanglement purification~\cite{bennett1996purification, deutsch1996quantum} and quantum error correction~\cite{fowler2010surface, javadi2017optimized} are promising means to suppress the effect of noise in the generation of entanglement. Both methods require the generation of many entangled pairs between the nodes, which are combined  to enable high-fidelity transfer of quantum information.% Standard entanglement generation protocols use photonic qubits to create entanglement between a pair of stationary qubits~\cite{simon2003robust, pompili2021realization, langenfeld2021quantum, lago2021telecom, delteil2016generation}. The generation is, however, probabilistic due to various loss channels of the photonic qubits such as the  exponential signal attenuation in optical fibers~\cite{}. In this case, one should restart the entanglement generation on the 'failed' channels and this will increase the required coherence time of the quantum memories.

\begin{figure}
    \centering
    \includegraphics[width=1.0\linewidth]{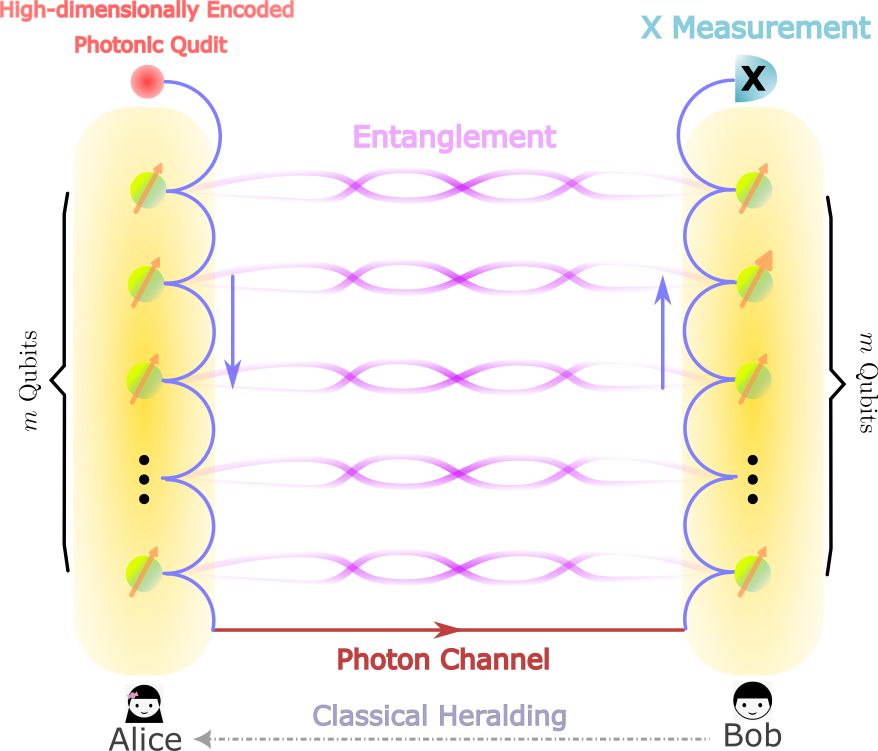}
    \caption{General schematic of previous photonic qudit-mediated entanglement generation protocols~\cite{zheng2022entanglement, xie2021quantum, piparo2019quantum, zhou2023parallel}. Alice and Bob have registers with $m$ qubits. A photonic qudit interacts with first Alice's and then Bob's qubit registers. Bob performs a high-dimensional $X$-basis measurement of the photonic qudit. Heralded on the detection of a photon, $m$ Bell pairs are generated simultaneously. Finally Bob classically communicates to Alice whether the protocol succeeded or not.}  \label{protocol}  
\end{figure}

Recently, several entanglement distribution protocols based on time-bin photonic qudit encoding were proposed for the simultaneous generation of multiple entangled pairs between network nodes~\cite{zheng2022entanglement, xie2021quantum, zhou2023parallel, piparo2019quantum}. In general, they follow the setup outlined in Fig.~\ref{protocol}. First, a single photonic qudit encoded in time-bins is entangled with Alice's qubit registers. The photon is transmitted to Bob, where the same entangling operation is performed. A successful run of the protocol is heralded by the detection of the photon via a high-dimensional $X$-basis measurement. If a photon is detected, all qubits in the registers of Alice and Bob are entangled. This contrasts with conventional schemes based on photonic qubits, where entangled pairs are generated independently of each other. This leads to conventional schemes having much more demanding requirements on the coherence times of the qubits for the generation of multiple entangled pairs. 

A significant experimental challenge of the schemes in Refs.~\cite{zheng2022entanglement, xie2021quantum, piparo2019quantum, zhou2023parallel} is the requirement of fast optical switches for routing of the photon to ensure the correct interaction with the qubits and to interfere the time-bins in the $X$-basis measurement of the photon. The schemes also require a single photon to interact with all qubits, which can cause correlated errors between the entangled pairs. The extent of correlated errors and their effects on key applications such as entanglement purification and teleportation-based error correction have not been studied in detail. 

In this paper, we propose a photonic qudit-mediated entanglement protocol that circumvents the use of optical switches, making it more accessible to the near-term hardware. We model and characterize the effect of correlated errors on the quality of the generated Bell pairs and apply optimized entanglement purification protocols~\cite{krastanov2019optimized} to distill high-fidelity entanglement. We also study the performance of teleportation-based quantum error correction, focusing on the [[5,1,3]] and [[4,2,2]] code \cite{knill2001scheme, knill2005quantum, namiki2016role, devitt2013quantum}. We show that the optimized purification protocols can remove the correlated errors efficiently while they have a more negative impact on the error correction codes.

\section{Switch-free Protocol}

\begin{figure*}
    \includegraphics[width=1.0\linewidth]{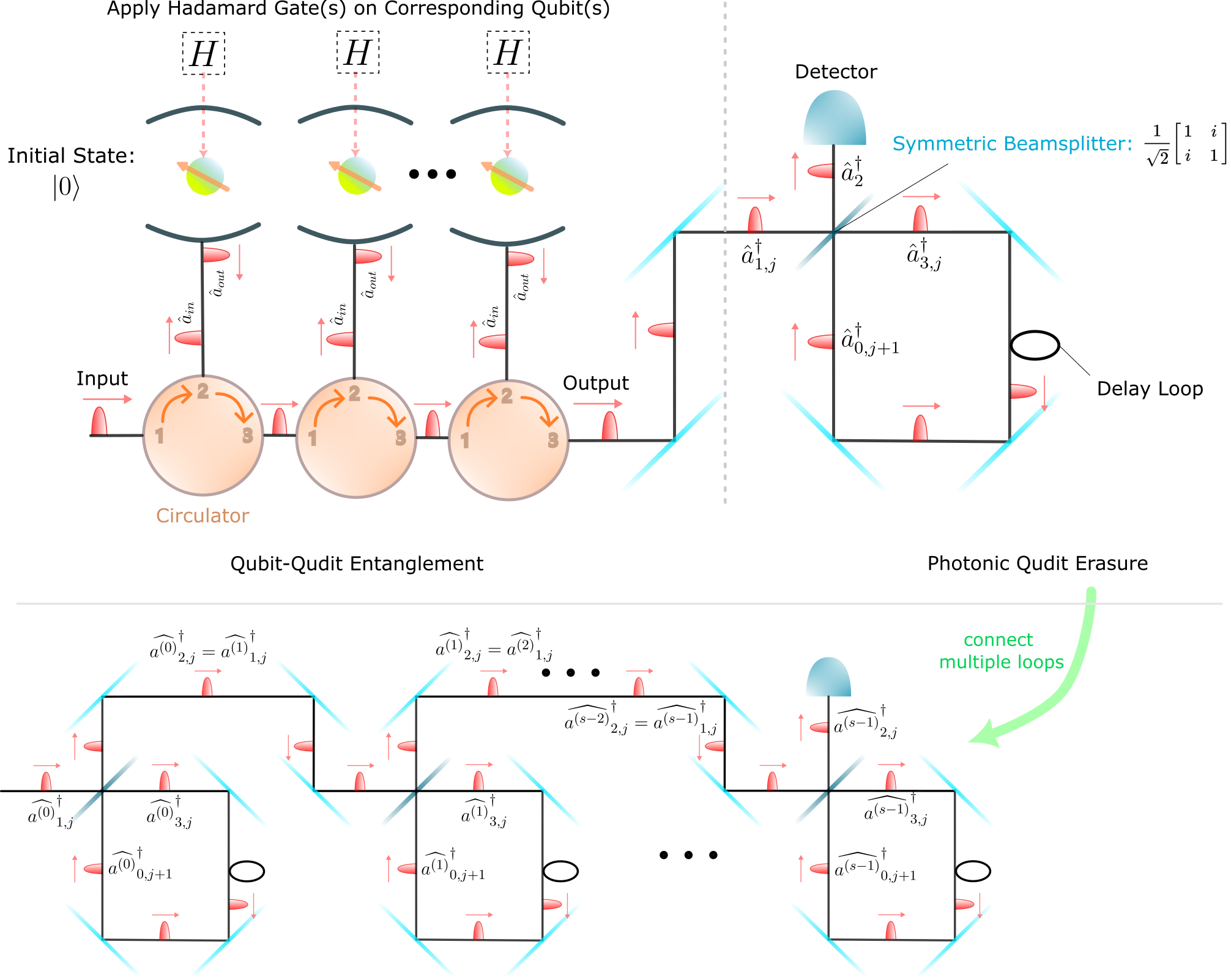}
    \caption{The main elements  of the switch-free qudit-mediated entangling protocol. On the top, left, we show the qubit-qudit entangling operation that Alice and Bob perform. The initial states of the qubits are $\ket{0}$ and Hadamard gate(s) are applied to specific qubits at each time-bin to ensure the correct qudit-qubit interactions. On the top, right, we show the final measurement of the photon, which erases the time-bin information and heralds successful entanglement generation. Upon arrival at the symmetric beam splitter, the photon is either reflected and detected or transmitted and delayed by exactly one time-bin. To boost the probability of a successful time-bin erasure, a concatenated series of beamsplitters and delay loops can be used as illustrated at the bottom of the figure.
    }  
    \label{compare}  
\end{figure*}

We show an overview of the switch-free implementation proposed in this paper in Fig.~\ref{compare}. We circumvent the need for optical switches in two ways. Firstly, each time-bin pulse generated by Alice will interact with all register qubits of both Alice and Bob through a single-sided cavity-mediated CZ-gate between the photon (logical states $\ket{0}/\ket{1}$ corresponding to absence/presence of the photon) and the qubits (logical states $\ket{0}/\ket{1}$ corresponding to a cavity non-coupled/coupled spin state). We further discuss the physics of the CZ-gate below. 

Through single qubit Hadamard gates, we can control whether the presence of the photon in a time-bin pulse will flip the state of the qubit or not. This is because the CZ-gate will only have an effect if the qubit state has a non-zero amplitude in the $\ket{1}$ state. Thus, there will be no effect if the qubit is in state $\ket{0}$, while if it is prepared in state $\ket{+}$, the scattering of the photon will flip the state to $\ket{-}$.  The  Hadamard gates will be applied at each time-bin following a binary logic as detailed below.

Secondly, in Ref.~\cite{zheng2022entanglement}, the final deterministic photon measurement also requires optical switches to map the time-bin encoding to a spatial encoding, allowing for interference of the different modes and erasure of the time-bin information. We show how the time-bin information can be erased probabilistically with a concatenated array of beamsplitters and delay lines. 

We now go through the steps of the protocol in more detail. First, Alice generates a time-bin photonic qudit with dimension $2^m$ where $m$ is the number of entangled pairs we want to generate. We model this generation assuming a cavity-assisted Raman scheme similar to the approach in Ref.~\cite{Knall2022}.  

We consider a 3-level system consisting of two stable ground states $\ket{g_0}$, $\ket{g_1}$ and one excited state $\ket{e_1}$. The transition $\ket{e_1}\leftrightarrow\ket{g_1}$ is coupled to an optical cavity for efficient photon extraction, while the transition $\ket{e_1}\leftrightarrow\ket{g_0}$ is driven by a pulsed laser. 

The system is initialized in state $\ket{g_0}$. At each time-bin, we drive with pulsed excitation such that there is a small probability that the system is transferred from the ground state $\ket{g_0}$ to $\ket{g_1}$ with the emission of a cavity photon.  By carefully tuning the amplitude of the driving pulses, a time-bin encoded photonic qudit of the following form will  be emitted in the ideal case:

\begin{equation}
\centering
\ket{\psi}_{ph}=\sum_{n=0}^{2^m-1}\alpha_{n}\ket{n}_{ph},
\label{photongen}
\end{equation}
where $\ket{n}_{ph}$ denotes a single photon in the $n$'th time-bin and vacuum in the rest. 

The amplitudes $\alpha_n\in\mathbb{R}^{+}$ can be tuned through the amplitude of the driving pulses. This becomes important later on as the probabilistic time-bin erasure step at the end of the entanglement generation protocol will affect the amplitude of the time-bins differently. This, as well as the general asymmetric loss in the setup, can be compensated by tuning the initial amplitudes. We discuss the exact form of $\alpha_n$ below.

Next, the generated photonic qudit interacts with the qubits in Alice's register. The register qubits of Alice and Bob are all initialized in state $\ket{0}$. Each time-bin pulse will interact with all register qubits through a cavity-mediated CZ-gate~\cite{duan2004scalable, bhaskar2020experimental}. This interaction ensures reflection of the photon from the cavity-spin system with (without) a $\pi$ phase shift if the spin is in an uncoupled (coupled) state to the cavity.

Suppose that the spin state is $\alpha \ket{0}+ \beta \ket{1}$, where only state $\ket{1}$ is coupled to an excited state via the cavity field. Reflecting a single photon in the first time-bin ($\ket{0}_{ph}$) will result in the transformation $\ket{0}_{ph}\otimes(\alpha \ket{0}+ \beta \ket{1})\to\ket{0}_{ph}\otimes(\alpha r_0\ket{0}+ \beta r_1 \ket{1})$. Ideally, $r_0=-1$ and $r_1 = 1$, which corresponds to a perfect CZ-gate up to a global phase. 

We can control whether the scattering of a photon will flip the state of a qubit through the application of Hadamard gates. This is done according to a binary encoding of the time-bin number. As an example, we consider the case of $m=3$. Starting from the initial state $\sum_{n=1}^{7}\alpha_{n}\ket{n}_{ph}\ket{000}$, the procedure would be:

{\small
\begin{itemize}
\setlength{\itemsep}{0pt}

    \item \textbf{At time-bin $\ket{0}_{ph}$:}
    {\footnotesize
    \subitem 1. no Hadamard gates applied
      \subitem 2. First time-bin pulse is reflected
      \subitem $\ket{0}_{ph}\ket{000}\rightarrow-\ket{0}_{ph}\ket{000}$
       \subitem 3. no Hadamard gates applied 
       \subitem $-\ket{0}_{ph}\ket{000}\rightarrow-\ket{0}_{ph}\ket{000}$
    }\\
    \textit{Resulting state:} \\$-\alpha_{0}\ket{0}_{ph}\ket{000}+\sum_{n=1}^{7}\alpha_{n}\ket{n}_{ph}\ket{000}$

    \item \textbf{At time-bin $\ket{1}_{ph}$:}
    {\footnotesize
       \subitem 1. Apply Hadamard on the first qubit
       \subitem 2. Second time-bin pulse is reflected
      \subitem $\ket{1}_{ph}\ket{+00}\rightarrow-\ket{1}_{ph}\ket{-00}$
       \subitem 3. Apply Hadamard on the first qubit
       }\\
    \textit{Resulting state:} \\$-\alpha_{0}\ket{0}_{ph}\ket{000}-\alpha_{1}\ket{1}_{ph}\ket{100}+\sum_{n=2}^{7}\alpha_{n}\ket{n}_{ph}\ket{000}$

    \item \textbf{At time-bin $\ket{2}_{ph}$:}
    {\footnotesize
       \subitem 1. Apply Hadamard on the second qubit
       \subitem 2. Third time-bin pulse is reflected
      \subitem $\ket{2}_{ph}\ket{0+0}\rightarrow-\ket{2}_{ph}\ket{0-0}$
       \subitem 3. Apply Hadamard on the second qubit
       }\\
    \textit{Resulting state:} \\$-\alpha_{0}\ket{0}_{ph}\ket{000}-\alpha_{1}\ket{1}_{ph}\ket{100}-\alpha_{2}\ket{2}_{ph}\ket{010}$\\
    $+\sum_{n=3}^{7}\alpha_{n}\ket{n}_{ph}\ket{000}$

    \item \textbf{At time-bin $\ket{3}_{ph}$:}
    {\footnotesize
       \subitem 1. Apply Hadamards on the first two qubits
       \subitem 2. Fourth time-bin pulse is reflected
      \subitem $\ket{3}_{ph}\ket{++0}\rightarrow-\ket{3}_{ph}\ket{--0}$
       \subitem 3. Apply Hadamards on the first two qubits
       }\\
    \textit{Resulting state:} \\$-\alpha_{0}\ket{0}_{ph}\ket{000}-\alpha_{1}\ket{1}_{ph}\ket{100}-\alpha_{2}\ket{2}_{ph}\ket{010}$\\
    $-\alpha_{3}\ket{3}_{ph}\ket{110}+\sum_{n=4}^{7}\alpha_{n}\ket{n}_{ph}\ket{000}$

    \item $\cdots$

\end{itemize}
}

Repeating this process for all time-bins, Alice's register qubits will be entangled with the photonic qudit in the following way:

\begin{equation}
\centering
\begin{split}
\ket{\psi}=&\alpha_0\ket{0}_{ph}\ket{000}+\alpha_1\ket{1}_{ph}\ket{100}+\\
&\alpha_2\ket{2}_{ph}\ket{010}+\alpha_3\ket{3}_{ph}\ket{110}+\\
&\alpha_4\ket{4}_{ph}\ket{001}+\alpha_5\ket{5}_{ph}\ket{101}+\\
&\alpha_6\ket{6}_{ph}\ket{011}+\alpha_7\ket{7}_{ph}\ket{111},
\end{split}
\end{equation}

after the scattering of the final (8th) time-bin up to a global phase. This can be extended straightforwardly to general $m$. Bob will perform the same procedure upon receiving the transmitted qudit from Alice. 

After the scattering of the final time-bin at Bob's side, the total state of the qudit and Alice and Bob's register qubits will be:

\begin{equation}
\centering
\begin{split}
\ket{\psi}=\sum_{n=0}^{2^m-1}\alpha_{n}\ket{n}_{ph}\ket{[\overline{n}]_2}_{A}\ket{[\overline{n}]_2}_{B},
\end{split}
\label{statebeforemeas}
\end{equation}
where $[\overline{n}]_2$ is the inverse of the binary formation of $i$ in digit size $m$. For example, when $m=3$, $[\overline{0}]_2\rightarrow{000}$, $[\overline{1}]_2\rightarrow{100}$, $[\overline{2}]_2\rightarrow{010}$, $[\overline{3}]_2\rightarrow{110}$, etc. 

Next, we need to unentangle the qudit from the spin qubits by measuring the qudit in a manner that erases the time-bin information. By measuring the photon we also herald that the protocol was successful.  

A single-loop, switch-free photon measurement approach is shown at the top, right half of Fig.~\ref{compare}. A symmetric beamsplitter will either transmit a time-bin pulse into a delay loop or reflect it to a single photon detector. The delay is tuned such that a delayed time-bin will interfere with the successive time-bin at the beamsplitter. 

To see how this can erase the time-bin information, we consider an example with only two time-bins. In this case, a detection at a time corresponding to the second time-bin will erase the information about whether the photon was initially in the first or second time-bin. Note, however, that the process is probabilistic since a detection at the first time-bin determines that the photon was initially in the first time bin.

This can be extended to interfering all time-bins through careful tuning of the initial qudit amplitudes, as we will show later. However, since the photon erasure is probabilistic, we can also connect multiple photon loops to increase the measurement success probability, as shown in the bottom of Fig.~\ref{compare}. Nevertheless, the connected delay loops lead to several interference effects not present in the single-loop case:

\begin{itemize}
\item Destructive interference. We find that due to destructive interference, certain detection times later than the $2^m$'th time-bin will not project into a superposition of $2^m$ spin states but only a subset of these. 
\item The tuning of $\alpha_n$ depends on the specific detection time and number of loops.  For the single loop, the tuning of the coefficients $\alpha_n$ does not depend on the exact detection time as long as this is later than or at the $2^m$'th time bin. However, in the connected loops, this is no not the case.
\item Phase correction problem. For the single loop, the phase due to the beam splitter transformation can always be corrected by single-qubit gates. For the connected loops, this is not always the case, depending on the specific detection time and the number of loops. 
\end{itemize}

We will now go through these setups in more detail and also refer the reader to Appendix~\ref{A4b} for additional information. For a setup with $s\geq1$ delay loops and a detection time of $u\geq2^m$ (in units of the time-bin duration), we define:

\begin{equation}
\boldsymbol{Y}(s,u,n)=\sum_{t=1}^{\min[s,u-n]}(-1)^{t+1}\binom{s}{t}\binom{u-n-1}{t-1}.
\end{equation}

%As we show in Appendix~\ref{A4b}, 
One has to choose $s$ and $u$ such that $\boldsymbol{Y}(s,u,n)$ have the same sign throughout all $n$, i.e.,

\begin{equation}
\begin{split}
&\boldsymbol{Y}(s,u,n)>0 \ \text{or} \ \boldsymbol{Y}(s,u,n)<0  \\
&\text{for} \ \forall n\in\{0,1,2,\dots,2^m-1\},
\end{split}
\end{equation}
to ensure that we can find suitable coefficients $\alpha_n$ such that the final qubit state following the photon detection is equivalent to $m$ copies of the Bell state $\ket{\Phi^+}=(\ket{0_A0_B}+\ket{1_A1_B})/\sqrt{2}$ up to single qubit gate corrections. The tuning of the coefficients is determined by

\begin{equation}
\centering
%\alpha_{n}^2=\frac{1}{\sum_{k=0}^{2^m-1}2^{n-k}\left|\frac{\boldsymbol{Y}(s,u,n)}{\boldsymbol{Y}(s,u,k)}\right|^2 },
\alpha_{n}^2=\left(\sum_{k=0}^{2^m-1}2^{n-k}\left|\frac{\boldsymbol{Y}(s,u,n)}{\boldsymbol{Y}(s,u,k)}\right|^2 \right)^{-1},
\label{alpha-connected}
\end{equation}

In this way, after the successful photon detection, we have the state:

\begin{equation}
\centering
\ket{\psi}=\frac{1}{2^{\frac{m}{2}}}\bigotimes_{n=0}^{m-1}(\ket{0_A0_B}+i^{-2^n}\ket{1_A1_B}).
\label{ABphase}
\end{equation}

where the extra phase $i^{-2^n}$ can be easily corrected by applying single qubit gates. The probability to detect the photon at time $u$ is:

\begin{equation}
\centering
%P_{succ}(s,u,m)=\frac{2^m}{\sum_{k=0}^{2^m-1}2^{s+u-k}\left|\frac{1}{\boldsymbol{Y}(s,u,k)}\right|^2}.
P_{succ}(s,u,m)=2^m\left(\sum_{k=0}^{2^m-1}2^{s+u-k}\left|\frac{1}{\boldsymbol{Y}(s,u,k)}\right|^2\right)^{-1}\!\!\!\!\!.
\end{equation}

It is important to note that, in the single loop case of $s=1$, there are no interference effects and measuring every time-bin after time-bin $2^m - 1$ leads to success. Therefore, one can find that in this case, the measurement always succeeds when $u\geq 2^m$, and the success probability for the single loop scenario should be the sum over all $P_{succ}(s=1,m,u\geq 2^m)$. 

\begin{figure}
    \includegraphics[width=1.0\linewidth]{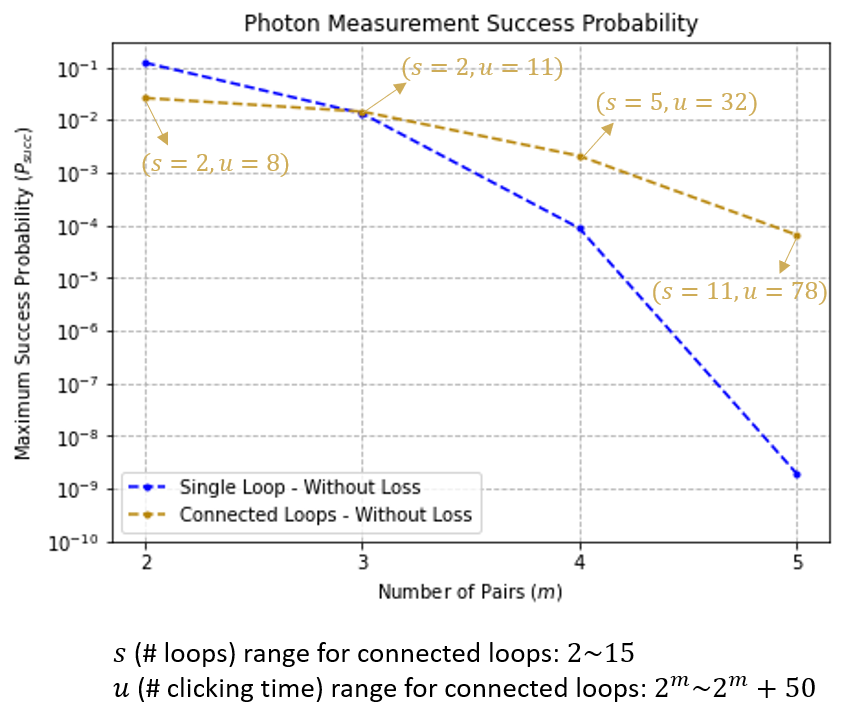}
    \caption{The photon measurement success probability under for a single loop and concatenated loops. For the single loop, the success probability is the sum over all $P_{succ}(s=1,m,u\geq 2^m)$. For the connected loops, the success probability shown here is the maximum probability within the ranges of $(s,u)$ shown above. The optimal $(s,u)$ for each number of entangled pairs is marked on the plot.}  
    \label{succ-prob-no-loss}  
\end{figure}

To illustrate the increase in probability by connecting more loops, we have plotted the success probabilities of measurements under both single loops and interconnected loops in Fig.~\ref{succ-prob-no-loss}. In this plot, we demonstrate that connecting loops can substantially enhance the success probability, especially for a larger number of Bell pairs, despite the fact that in the single-loop cases, all instances where $u\geq 2^m$ can be considered as successful. %We limit the number of connected loops to $s\geq2$ and the detection time to $u\geq2^m$ within specific ranges. We identify the optimal set of $(s,u)$ that yields the maximum success probability. This confinement is imposed due to the potential increase in probabilities of additional noise and decoherence on the photonic qudit when adding more connected loops or prolonging the waiting time. It is worth noting that introducing additional values for $s$ and $u$ only results in a trivial increase in the success probability.

\section{\label{errorcor}Error Analysis and Error Correlations}

\begin{figure*}
    \includegraphics[width=0.9\linewidth]{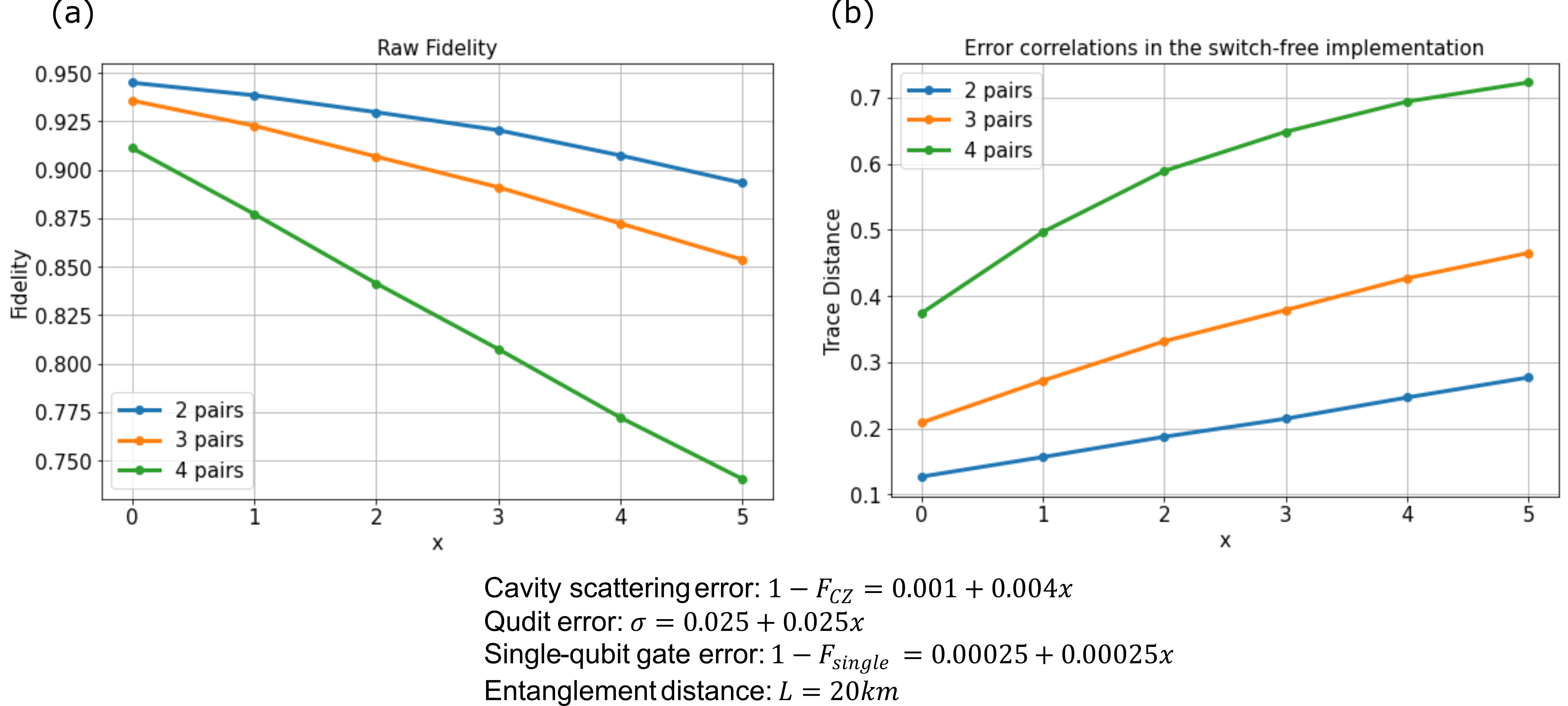}
    \caption{(a) Raw fidelity of the generated Bell pairs using the switch-free qudit protocol. (b) Error correlations under the specified error environment. One can notice that in (a) and (b), the raw fidelity will decrease and the error correlations will grow when increasing errors and the number of generated pairs.}  
    \label{correlationresult}  
\end{figure*}

So far, we have not considered the effect of noise and general experimental imperfections. These will affect both the success probability of the protocol and the fidelity of the generated Bell pairs. To analyze this, we model the effect of the following possible imperfections on our protocol: 1) imperfections in the generation of the photonic qudit due to phase and amplitude fluctuations of the laser drive; 2) imperfect cavity scattering in the spin-photon CZ-gate; 3) single-qubit gate errors; 4) spin qubit decoherence; and 5) photon losses. 

Photon loss is different from the other imperfections as we herald on the detection of a photon. %The fidelity only decreases when a photon is lost but the photon detector clicks due to dark counts. 
In our analysis, we assume dark counts to be negligible and that the loss is accurately characterized such that we can compensate the effect on the fidelity of the generated entanglement by tuning the initial amplitudes of the photonic qudit (see Appendix~\ref{A4b} for details). Photon loss will therefore only affect the success probability of the protocol.  

The other imperfections will decrease the fidelity of the generated Bell pairs since they are not heralded. Moreover, they can lead to correlated errors, which means that some errors may be propagated to multiple generated Bell pairs. This can occur since all register qubits are entangled with one single photon and any imperfections of the photon and operations may lead to correlated errors. The details of our error modeling are presented in Appendix~\ref{A}. The decrease in the average fidelity of the generated Bell pairs is shown in Fig.~\ref{correlationresult} (a) for a specific range of error parameters.

To quantify the amount of error correlation between the Bell pairs, we use the following approach. 
Suppose the probabilities that a Pauli error happens on qubits $a$ and $b$ are $\epsilon_a(\ll1)$ and $\epsilon_b(\ll1)$, respectively, while the probability that a Pauli error happens on both qubits $a$ and $b$ is $\epsilon_{ab}$. If $\epsilon_{ab} = \epsilon_a \epsilon_b$, the errors on qubits $a$ and $b$ are independent and we call them "uncorrelated" throughout this paper, while if $\epsilon_{ab} \neq \epsilon_a \epsilon_b$ holds, the errors are dependent and we refer to them as "correlated" errors.
%\footnote{\textcolor{red}{@Liubov, could you please write something about uncorrelated (independent) / correlated (dependent)? You of course understand them better than me. I still prefer using terms '(un)correlated' for reasons of consistency, since they are compatible and intensively used in this paper and also Yunzhe's paper (https://journals.aps.org/prxquantum/abstract/10.1103/PRXQuantum.3.040319). But you are absolutely right, we should elaborate this here for not making confusions to readers. Could you please help me with that? Thanks!}}. 
%\liubov{don't write correlated, write dependent. it can make a confusion, since if  $\epsilon_{ab} \neq \epsilon_a \epsilon_b$ holds the correlation is not zero in general but only for Gaussian rv. I know you meant different correlation but better to use dependent further. Also further is it not better to use $\rho_d$ and $\rho_i$ for dependent and independent? Its shorter}

In general, an independent error model is of the form
\begin{equation}
\centering
\rho_{\text{uncorrelated}}=\Lambda_{2m}(\dots(\Lambda_{3}(\Lambda_{2}(\Lambda_{1}(\rho_{perfect})))),
\label{unco}
\end{equation}

where $\Lambda$ denotes the corresponding Pauli error channel $(I, X, Y, Z)$ on each register qubit ($2m$ in total). In this case, multi-qubit errors happen with very low probability since $\epsilon_{ab} \ll \epsilon_a$,$\epsilon_b$. However, if $\epsilon_{ab} \neq \epsilon_a \epsilon_b$, the errors among the pairs are correlated. We quantify the amount of error correlations for the qudit-mediated entanglement generation in the following way. 

Let $\rho_{\text{correlated}}$ be the output density matrix of the qubits after a successful entanglement generation of the qudit protocol for a specific choice of error parameters. The fidelity between this state and the perfect output state (collection of $m$ Bell pairs) is denoted by $F(\rho_{\text{correlated}})$. Similarly, the fidelity between the perfect state and the output of the uncorrelated error channel in \eqref{unco} is $F(\rho_{\text{uncorrelated}})$. We then let $T_{\min}=\mathop{\min}\limits_{\Lambda}  T(\rho_{\text{uncorrelated}}, \rho_{\text{correlated}})$ s.t. $|F(\rho_{\text{correlated}})-F(\rho_{\text{uncorrelated}})|\leq\epsilon$ for sufficiently low $\epsilon$ (we choose $\epsilon=10^{-4}$ in this work) quantify how large the error correlation is.

%{\small
%\begin{enumerate}
%\setlength{\itemsep}{0pt}
%    \item For a specific choice of error-parameters, we find the output density matrix $\rho_{correlated}$ of the spins after a successful entanglement generation of the qudit protocol.
%    \item We calculate the average fidelity compared with a collection of perfect Bell pairs $F(\rho_{correlated}, \rho_{perfect})$. 
%    \item We numerically find an independent error model that gives an output density matrix (see Eq.~(\ref{unco})) such that $|F(\rho_{uncorrelated}, \rho_{perfect})-F(\rho_{correlated}, \rho_{perfect})|\leq\epsilon$ . Note that there are multiple error channels that satisfy this. 
%    \item Finally, we minimize the trace distance between the correlated and uncorrelated density matrices wrt to the choice of uncorrelated error model $\Lambda$, under the constraint that the fidelity should be almost the same within $\epsilon$. $T_{min}=\mathop{\min}\limits_{\Lambda}  T(\rho_{uncorrelated}, \rho_{correlated})$.
%    \item The minimum trace distance $T_{min}$ denotes how large the error correlation is.
%\end{enumerate}
%}

\begin{figure*}
    \includegraphics[width=0.9\linewidth]{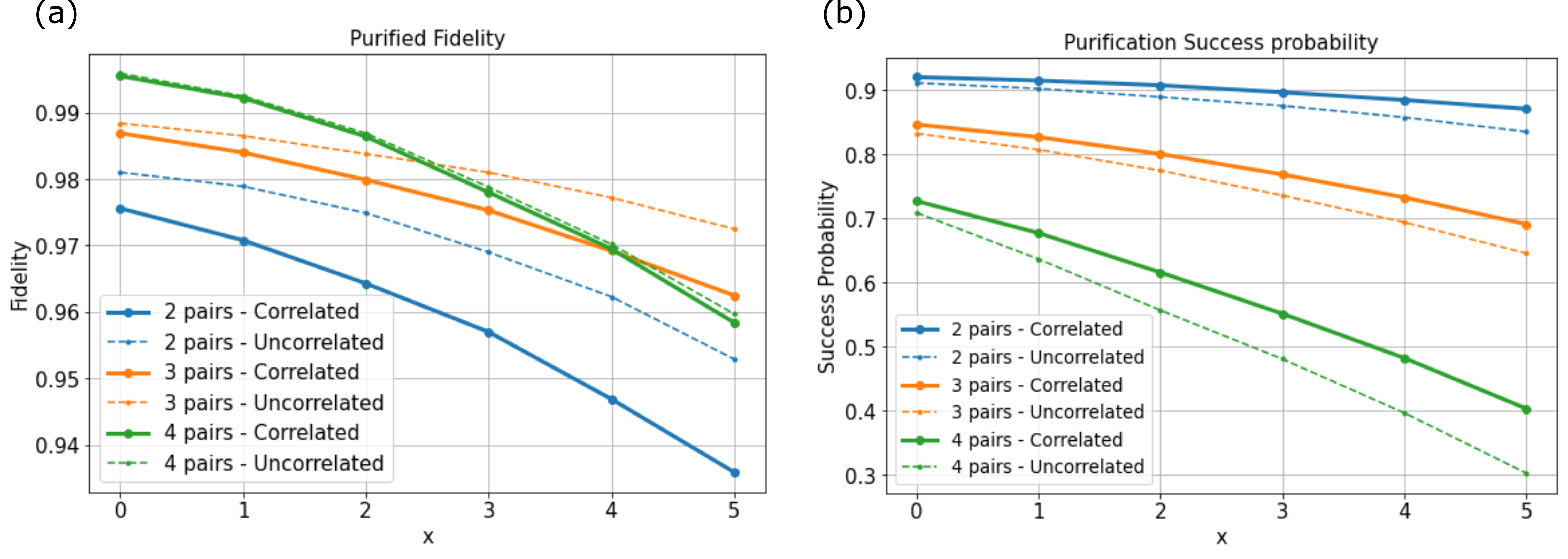}
    \caption{(a) Purified fidelity for error-correlated and error-uncorrelated pairs, and (b) purification success probability under the specified error environment shown below. The error range is the same as in Fig.~\ref{correlationresult}. For the purification plot (a) and (b), one can see an obvious fidelity increase after the purification (the raw pair fidelity has been shown in Fig.~\ref{correlationresult} (a)). The error range shown in the plot is the same as in Fig.~\ref{correlationresult}. % For smaller errors, purification with more pairs will perform with higher purified fidelity, but the success probability will decrease as well. In (a), we also include the case of the error-uncorrelated pairs, and we can see that the correlated errors can be mitigated efficiently due to the protocol's optimization features.  Note that we assume perfect operations in the purification circuits.
    }  
    \label{puriresult}  
\end{figure*}

Fig.~\ref{correlationresult} (b) shows how the amount of error correlations in the qudit protocol increases as the general errors increase. Furthermore, we see how the amount of correlation increases with the number of pairs. In the following two sections, we consider two strategies for mitigating the effect of the errors and obtain high-fidelity Bell pairs: entanglement purification and quantum error correction.

\section{Purification Performance}

Entanglement purification is a method that aims to produce one (or possibly more) high-fidelity entangled pair(s) from a collection of noisy pairs \cite{bennett1996purification, deutsch1996quantum,dur2007entanglement}. It is a key element to reducing noise in first-generation quantum repeaters \cite{muralidharan2016optimal}. 

To search for optimal purification protocols, we use the evolutionary algorithm from Ref.~\cite{krastanov2019optimized}, which enables $m$-to-one purification for any number of generated noisy Bell pairs $m$. 
%The basic idea of this algorithm is to choose the optimized Clifford building fragments to construct the local purification operations. The operation fragments are designed to permute the error terms in the density matrix such that the most dominant errors are detected through comparison of measurement outcomes between Alice and Bob. This genetic purification protocol is shown in detail in Ref.~\cite{krastanov2019optimized} and Appendix~\ref{B}.
The performance of the optimized purification circuits is shown in Fig.~\ref{puriresult} (a) and (b). We see that the purification protocol successfully produces a higher fidelity entangled pair from the collection of noisy pairs. %Also, for smaller errors, generating more pairs and then applying $m$-to-1 purification from them will very possibly provide higher fidelity. But with a larger number of generated pairs, the purification success probability is decreasing as well since more (anti-)coincidence measurement conditions are required.

\begin{figure*}
    \includegraphics[width=1.0\linewidth]{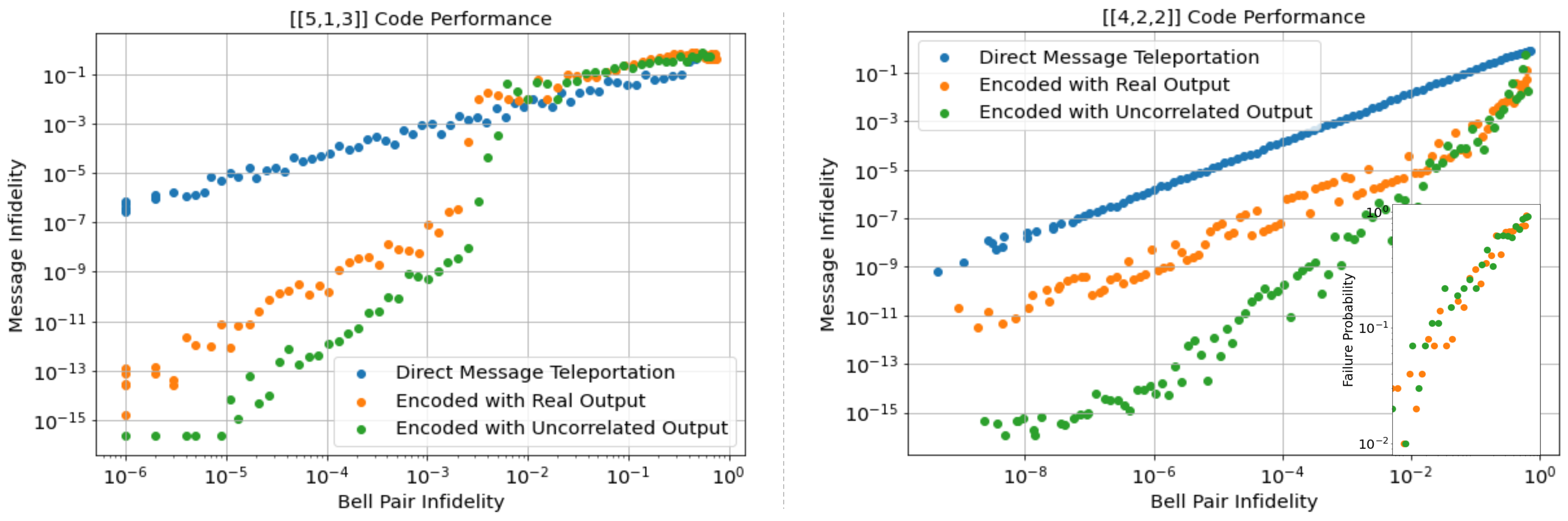}
    \caption{The performance of teleportation-based [[5,1,3]] and [[4,2,2]] code under the switch-free implementation. The $x$ and $y$ axes refer to the Bell pair infidelity and the transmitted message infidelity, respectively. The blue dots denote the data points with direct message teleportation i.e. without encoding and using a single Bell pair. The orange and green dots denote the data with the correction code, teleported by multiple Bell pairs with correlated (real) and uncorrelated error models, respectively. The rightmost plot for each simulating code indicates the data with lower correlated errors. The inset plot depicts the failure probability of the [[4,2,2]] code, indicating the likelihood of detecting errors. We assume perfect encoding and decoding operations.}  
    \label{513422}  
\end{figure*}

We also investigate how well this optimized purification protocol can mitigate the error correlations. For each error data point, we find the matching uncorrelated error channel following the approach outlined in Sec.~\ref{errorcor}. We find the optimal purification protocol for the uncorrelated error channel and compare it with the performance of the optimized purification of the qudit-mediated entangled state. The result is shown in Fig.~\ref{puriresult} (a). We see that for the purification of a few pairs, the purification of the qudit-mediated entanglement performs slightly worse. However, the performance difference decreases as the number of pairs increases since the purification circuit has more freedom to target the dominant error terms. Thus, it seems that optimizing the purification circuit allows to target the dominant errors regardless of whether they are correlated or not. Note that we assume perfect operations in the purification circuit in Fig.~\ref{puriresult} (a) and (b) to focus on the ability to target the errors from the entanglement generation procedure.

\section{Teleportation-based Quantum Error Correction}

Teleportation-based quantum error correction is another method to correct for noisy connections in a quantum network, as exploited in so-called second generation quantum repeaters~\cite{muralidharan2016optimal, fowler2010surface, javadi2017optimized} and modular quantum computing~\cite{monroe2014large}.
%A $[[n,k,d]]$ quantum error correcting code means that one can encode $k$ logical qubits into $n$ physical qubits with a code distance of $d$. The code distance is the minimum number of operations on the physical qubits to perform a logical operation. 
In the teleportation-based setting, Alice wishes to teleport an encoded message to Bob via multi-qubit quantum teleportation~\cite{zhang2005many, li2016quantum} requiring the simultaneous availability of multiple entangled pairs~\cite{knill2001scheme, knill2005quantum, namiki2016role, devitt2013quantum}. 

We consider two simple codes: the [[5,1,3]] and the [[4,2,2]] codes requiring 5 and 4 Bell pairs, respectively, for teleportation. The former is the smallest code that can perform quantum error correction \cite{laflamme1996perfect}, and the latter only detects the errors~\cite{grassl1997codes}. The Bell pairs are generated with the qudit protocol, and we assume perfect operations in the encoding and decoding of the message to focus on the ability of the code to correct errors from the noisy Bell pairs. 

The simulation results are shown in Fig.~\ref{513422}, where we compare the fidelity of the encoded transmission to the fidelity of teleporting unencoded qubits. We sample the qubit states (1-qubit teleportation for [[5,1,3]] code and 2-qubit for [[4,2,2]]) to be teleported at random, and each data point denotes the average result over all sampled random cases. Notably, for [[5,1,3]] code, there is a threshold above which the encoding performs worse than the unencoded transmission. This threshold is for errors $\sim10^{-3}$. For [[4,2,2]] code, however, teleporting the perfect code will always win if considering the transmitted message fidelity only. However, since the [[4,2,2]] code cannot correct the errors but only detect them, the transmitted message will fail if an error is detected, resulting in a non-deterministic transfer. 

We also investigate the effect of correlated errors on the performance of the error correction, following the same approach as for entanglement purification.  We see from Fig.~\ref{513422} that the codes perform worse on the error-correlated states. In contrast to the purification circuit, these codes are also not designed for correlated errors, which might be a subject for further study.   

\section{Conclusion}

In this paper, we have proposed a novel implementation of qudit-mediated entanglement generation protocols that completely circumvents the use of optical switches. Notably, we show how probabilistic erasure of the photon time-bin information can be implemented with passive linear optics elements and how the success probability can be increased through the concatenation of beam splitters and delay loops.  

In addition, we have simulated the amount of error correlation that the qudit protocol introduces between the simultaneously generated Bell pairs and its impact on key applications such as entanglement purification and quantum error correction. We found that significant error correlations are introduced, but through optimization of the purification circuit~\cite{krastanov2019optimized}, the errors can still be efficiently targeted. For the two quantum error correcting codes, the [[5,1,3]] and [[4,2,2]] codes that we considered for teleportation negatively impact the performance. 

Given the significantly reduced memory requirements for the spin qubits~\cite{zheng2022entanglement}, this protocol can be used to boost the performance of current quantum hardware for quantum networking. In particular, quantum repeaters, where the distillation of high-fidelity Bell pairs through entanglement purification is a key requisite, necessitate the availability of multiple Bell pairs at the same point in time. Importantly, we have shown that high-fidelity entanglement can indeed be distilled from the simultaneously generated Bell pairs in the qudit protocol, even in the presence of correlated errors. 

%Notice that both entanglement purification and teleportation-based error correction (detection) code can boost the capability of the quantum networks. However, entanglement purification requires two-way classical heralding to determine the success of the purification, but teleportation-based error correction (detection) code only needs one-way classical information transmission during the quantum teleportations. Therefore, the latter suffers from shorter decoherence time due to its shorter heralding waiting time. Nonetheless, the purification protocol is widely adaptable due to its low requirement for the initially generated noisy pairs.

\begin{acknowledgments}
We thank Yunzhe Zheng (TUDelft) and Hemant Sharma (TUDelft) for the supportive discussions about the qudit-mediated entanglement distributions and cavity scatterings. We thank Stefan Krastanov (UMass Amherst) for the useful comments and inputs on the genetically-optimized purification protocol. We acknowledge funding from the NWO Gravitation Program Quantum Software Consortium (Project QSC No. 024.003.037). J.B. acknowledges support from The AWS Quantum Discovery Fund at the Harvard Quantum Initiative. X.L. acknowledges support received by the Dutch National Growth Fund (NGF), as part of the Quantum Delta NL programme. L.M. partly carried out her work in the framework of the Russian Quantum Technologies Roadmap.
L.M. was  supported by the Netherlands Organisation for Scientific Research (NWO/OCW), as part of the Quantum Software Consortium program (project  QSC No. 024.003.037 / 3368).
\end{acknowledgments}

\appendix

\section{\label{A}Error Models}

Here we discuss the analytical error estimations for the switch-free, qudit mediated entanglement generation protocol.

\subsection{\label{A1}Photonic Qudit Generation}

\begin{figure}
    \centering
    \includegraphics[width=0.75\linewidth]{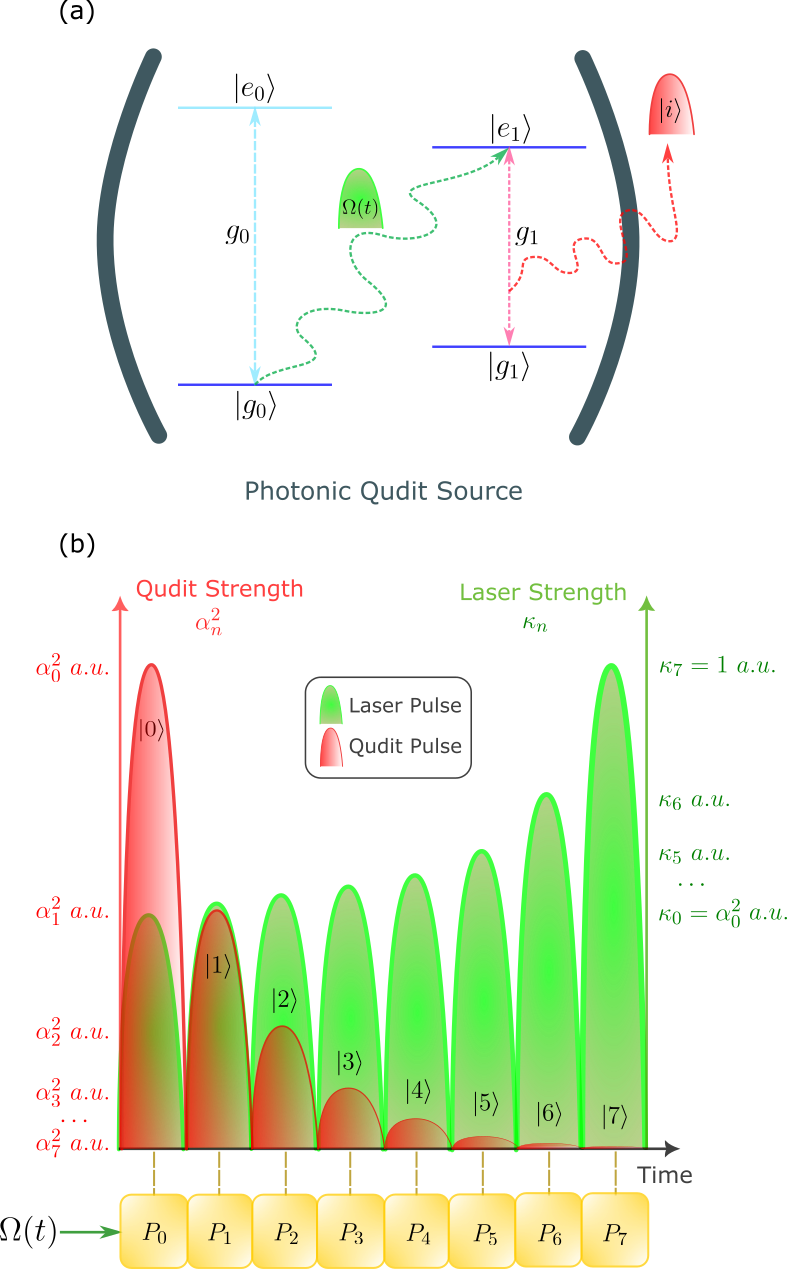}
    \caption{(a) Schematic of the photonic qudit. The laser pulses drives a cavity-asssited Raman transition from the initial state $\ket{g_0}$ to $\ket{g_1}$ with the emission of a photon. (b) Example of how varying the amplitude of the laser pulses affect the amplitude of the resulting qudit state.%The error bars denote that the amplitude errors on qudit pulses are also not constant, since they are depending on the magnitude between the strengths of the laser driving and qudit pulse. Note that in the above example, we input the corresponding $\alpha_n$ that the final measurement system has only one single photon loop and no losses are considered. The pulse strength will change correspondingly if multiple photon loops are connected, and/or the photon losses are considered.
    } 
    \label{quditgenfig}
\end{figure}

At the beginning of the switch-free implementation, Alice should generate a photonic time-bin qudit with $2^m$ pulses. We consider a Raman scheme for photonic qudit generation, which has been experimentally realized in quantum dot \cite{lee2018controllable} and silicon-vacancy diamond system \cite{knall2022efficient}. The setup is shown in Fig.~\ref{quditgenfig} (a). We initialize the emitter in state $\ket{g_0}$. At each time-bin, a laser pulse drives with a tunable probability the transition from $\ket{g_0}$ to $\ket{g_1}$ with the emission of a cavity photon.

Here, we illustrate an example. Suppose we would like to generate an 8-time-bin photonic qudit illustrated in Eqs.~(\ref{photongen}) (where $m=3$):

\begin{widetext}
\begin{itemize}
\setlength{\itemsep}{0pt}
    \item \textbf{At time-bin $\ket{0}_{ph}$}, the pulse $P_0$ drives $\ket{g_0}$ to $\ket{g_1}$ with the emission of $\ket{0}_{ph}$ resulting in the state:
    \begin{equation}
    \centering
    \begin{split}
    \ket{g_0}&\rightarrow\sqrt{1-\alpha_0^2}\ket{g_0}\ket{vac}_{ph}+\alpha_0\ket{g_1}\ket{0}_{ph}
    \end{split}
    \end{equation}
    
    \item \textbf{At time-bin $\ket{1}_{ph}$}, the pulse $P_1$ drives again $\ket{g_0}$ to $\ket{g_1}$ with the emission of $\ket{1}_{ph}$ such that:
    \begin{equation}
    \centering
    \sqrt{1-\alpha_0^2}\ket{g_0}\ket{vac}_{ph}+\alpha_0\ket{g_1}\ket{0}_{ph}\rightarrow\sqrt{1-\alpha_0^2-\alpha_1^2}\ket{g_0}\ket{vac}_{ph}+\ket{g_1}(\alpha_0\ket{0}_{ph}+\alpha_1\ket{1}_{ph})
    \end{equation}    
    
    \item \textbf{At time-bin $\ket{2}_{ph}$}, the pulse $P_2$ drives $\ket{g_0}$ to $\ket{g_1}$ with the emission of $\ket{2}_{ph}$ such that:
    \begin{equation}
    \centering
    \sqrt{1-\alpha_0^2-\alpha_1^2}\ket{g_0}\ket{vac}_{ph}+\ket{g_1}\sum_{n=0}^{1}\alpha_n\ket{n}_{ph}\rightarrow\sqrt{1-\left(\sum_{n=0}^{2}\alpha_n^2\right)}\ket{g_0}\ket{vac}_{ph}+\ket{g_1}\sum_{n=0}^{2}\alpha_n\ket{n}_{ph}
    \end{equation}  
    
    \item $\cdots$
    
\end{itemize}
\end{widetext}

Continuing the procedures until the final time-bin $\ket{7}_{ph}$, we will have removed all the amplitude of the $\ket{g_0}$ state such that the emitter is in state $\ket{g_1}$ and the photonic state is:

\begin{equation}
\centering
\ket{\psi}_{ph}=\sum_{i=0}^{7}\alpha_n\ket{n}_{ph},
\end{equation}

where  $\sum_{n=0}^{2^m-1}\alpha_n^2=1$. This procedure can readily be extented to any qudit dimension.

Note that at each time-bin, different laser driving strengths should be applied to generate each time-bin pulse, as shown in Fig.~\ref{quditgenfig} (b). For example, as shown for the 8-time-bin case in figure \ref{quditgenfig}: $P_0=\alpha_0^2$, $P_1=\frac{\alpha_1^2}{1-\alpha_0^2}$, $P_2=\frac{\alpha_2^2}{1-\alpha_0^2-\alpha_1^2}$,\dots, $P_7=\frac{\alpha_7^2}{1-\sum_{k=0}^{6}\alpha_k^2}$. For a general $2^m$-time-bin case, we have:

\begin{equation}
\centering
P_{0}=\alpha_0^2,
\end{equation}

and

\begin{equation}
\centering
P_{n}=\frac{\alpha_n^2}{1-\sum_{k=0}^{n-1}\alpha_k^2}, \ \text{for} \ 1\leqslant n\leqslant2^m-1.
\end{equation}

Imperfect cavity coupling will predominantly lead to loss of the photon or dephasing of the qudit state through spontaneous decay from the excited state $\ket{e_1}$ followed by re-emission. The first can be absorbed in the general transmission probability of the photon in the scheme. In addition, phase and amplitude fluctuations of the laser drive can also lead to dephasing and modulation of the amplitudes of the desired qudit state in addition to dephasing of the spin states. 

All the errors considered above lead to dephasing and modulation of the amplitudes of the desired qudit state. We choose a simplified model where we  model these imperfactions as random Gaussian fluctuations applied to the laser drive since this will lead to the same effect. The output qudit state is thus modelled as:%Note that for amplitude errors, due to the magnitude differences between the laser drive and qudit pulse at certain time-bins, the errors on qudit pulses may vary from the ones in laser drivings. Suppose the amplitude error on a certain laser driving is $\zeta_n$, then the amplitude error on the corresponding qudit pulse will be $\frac{\alpha_n^2}{P_n}\zeta_n$ (as shown in the error bars of Fig.~ \ref{quditgenfig} (b)). We also include a Gaussian phase error $\theta_n$ on the laser driving, and finally, we have (normalization omitted):

\begin{equation}
\label{noisyqudit}
\centering
\begin{split}
\ket{\psi}_{ph}&=\sum_{n=0}^{2^m-1}\sqrt{1+\frac{\zeta_{n}}{P_{n}}}\alpha_n e^{i\theta_{n}}\ket{n}_{ph}\\
&\approx\sum_{n=0}^{2^m-1}\left(1+\frac{\zeta_n}{2P_n}\right)\alpha_n e^{i\theta_{n}}\ket{n}_{ph}.
\end{split}
\end{equation}

For simplicity, we consider $\theta_n\sim N(0,\sigma^2)$ and $\zeta_n\sim N(0, \frac{1}{\sum_{k=0}^{2^m-1}\alpha_k^4P_k^{-2}}\sigma^2)$. This is to make sure that the total amplitude error gives: $\sum_{n=0}^{2^m-1}\frac{\alpha_n^2}{P_n}\zeta_n\sim N(0,\sigma^2)$.

\begin{figure}
    \includegraphics[width=0.7\linewidth]{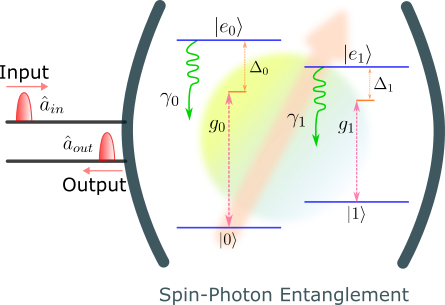}
    \caption{The structure of spin-photon entanglement system in a 4-level cavity \cite{hemant, tiurev2021fidelity, yang2004simplified, zheng2022entanglement} is shown. $\hat{a}_{in}$ and $\hat{a}_{out}$ are the photonic input and output of the cavity system. $\ket{e_0}$ and $\ket{e_1}$ denote the corresponding excited states of $\ket{0}$ and $\ket{1}$. $g_0$ and $g_1$ are the transition couplings. $\Delta_0$ and $\Delta_1$ are the detunings between the transition coupling and the cavity mode. $\gamma_0$ and $\gamma_1$ denote the spontaneous emissions of $e_0\rightarrow\ket{0}$ and $e_1\rightarrow\ket{1}$, respectively.} 
    \label{cav}  
\end{figure}

\subsection{\label{A3}Cavity Scattering}

In the main text, we discussed that after scattering of a photon, the qubit state $\alpha \ket{0}+ \beta \ket{1}$ will become $\alpha r_0\ket{0}+ \beta r_1\ket{1}$, and ideally, $r_0 \approx -1$ and $r_1 \approx 1$, which forms a perfect control-Z gate. In the non-ideal case, however, one may consider the four-level cavity system with two ground states $\ket{0}$, $\ket{1}$ and their excited states $\ket{e_0}$, $\ket{e_1}$ shown in Fig.~\ref{cav}~\cite{hemant, tiurev2021fidelity, yang2004simplified, zheng2022entanglement}, as this is a good model for realistic hardware such as quantum dots and group IV diamond vacancy systems. 

Here, we briefly introduce the model used in this work. One may refer to Ref.~\cite{zheng2022entanglement} for more details. We denote $\hat{c}$ the annihilation operator of the cavity field and $\Delta_0$ ($\Delta_1$) is the detuning between the transitions $\ket{0}\leftrightarrow\ket{e_0}$ ($\ket{1}\leftrightarrow\ket{e_1}$) and the cavity mode. The coupling rate of the two transitions is $g_0$ and $g_1$, respectively. The Hamiltonian $H$ of the system is:

\begin{equation}
\centering
\begin{split}
H=&\Delta_0\ket{e_0}\bra{e_0}+\Delta_1\ket{e_1}\bra{e_1} \\
&+(g_0\ket{e_0}\bra{0}\hat{c}+h.c.)+(g_1\ket{e_1}\bra{1}\hat{c}+h.c.),
\end{split}
\end{equation}

Considering the spin-photon interaction in the weak-driving regime, one can derive the reflection coefficients $r_0$ ($r_1$) for the state $\ket{0}$ ($\ket{1}$): 

\begin{equation}
\centering
\begin{split}
&r_0=1-\frac{K_{in}/K}{-\frac{i\omega}{K}+\frac{1}{2}+\frac{C_0}{-\frac{i}{\gamma_0}(\omega+\Delta_0)+\frac{1}{2}}}, \\
&r_1=1-\frac{K_{in}/K}{-\frac{i\omega}{K}+\frac{1}{2}+\frac{C_1}{-\frac{i}{\gamma_1}(\omega+\Delta_1)+\frac{1}{2}}},
\end{split}
\end{equation}

where $K_{in}$ is the decay rate of the cavity field into the collected mode and $K_{loss}$ is the cavity loss rate. Thus, the total cavity decay rate is $K=K_{in}+K_{loss}$. We also define the cooperativities $C_0$ and $C_1$ as:

\begin{equation}
\centering
C_0=\frac{|g_0|^2}{K\gamma_0}, \ \ \ 
C_1=\frac{|g_1|^2}{K\gamma_1}.
\end{equation}

For example, if the atomic qubit state is in a perfect $\ket{+}=\frac{1}{\sqrt{2}}(\ket{0}+\ket{1})$, after the cavity scattering, we get the following state  (normalization omitted):

\begin{equation}
\centering
\ket{n}_{ph}\otimes\frac{1}{\sqrt{2}}(\ket{0}+\ket{1})\xrightarrow{CS}\ket{n}_{ph}\otimes\frac{1}{\sqrt{2}}(r_0\ket{0}+r_1\ket{1}),
\end{equation}

where $n\in[0,2^m-1]$. Ideally, $C\gg 1$ and $K_{in}\approx K$. Then we have:

\begin{equation}
\centering
r_0\approx -1, \ r_1\approx 1.
\end{equation}

Therefore, in the ideal cases, this interaction will form a control-Z gate applying to the atomic qubit in the cavity.

\subsection{\label{A4b}Photon Measurement}

\subsubsection{Single Delay Loop}

We first calculate the case where there is one delay loop in the time-bin erasure. The right-hand side of Fig.~\ref{compare} (b) shows an overview of this setup. We write the state \eqref{statebeforemeas} as follows

\begin{equation}
\centering
\begin{split}
\ket{\psi}&=\frac{1}{2^{\frac{m}{2}}}\sum_{n=0}^{2^m-1}\alpha_{n}\ket{n}_{ph}S_n \\
&=\frac{1}{2^{\frac{m}{2}}}\sum_{n=0}^{2^m-1}\hat{a}^{\dagger}_{1,n}\ket{vac}_{ph,n}\alpha_{n}S_{n}.
\end{split}
\label{nnnnn}
\end{equation}

Here, index $n$ denotes the time-bin and $S_n=\ket{[\bar{n}]_2}_{A} \ket{[\bar{n}]_2}_{B}$. We denote $i$ as the imaginary unit. As shown in Fig.~\ref{compare}, the indices $0$, $1$, $2$, and $3$ denote the operator positions at the four ports of the symmetric beamsplitter $\tau$:

\begin{equation}
\centering
\tau=\frac{1}{\sqrt{2}}\left[ \begin{matrix}
   1 & i  \\
   i & 1  \\
\end{matrix} \right].
\end{equation}

Therefore, for the four ports of the symmetric beamsplitter $\tau$, we have:

\begin{equation}
\centering
\begin{split}
&\hat{a}^{\dagger}_{0,j}=\frac{1}{\sqrt{2}}(\hat{a}^{\dagger}_{2,j}+i\hat{a}^{\dagger}_{3,j}), \\
&\hat{a}^{\dagger}_{1,j}=\frac{1}{\sqrt{2}}(i\hat{a}^{\dagger}_{2,j}+\hat{a}^{\dagger}_{3,j}).
\end{split}
\end{equation}

Here $j$ denotes the time-bin. The delay line between $\hat{a}^{\dagger}_{3,j}$ and $\hat{a}^{\dagger}_{0,j+1}$ moves a pulse in time-bin $j$ to time-bin $j + 1$. Ideally, this achieves $\hat{a}^{\dagger}_{3,j}\rightarrow\hat{a}^{\dagger}_{0,j+1}$. However, there can be an additional probability of losing photons in the delay loop which will affect the amplitude of the pulses. We model this with a beam splitter with the following transfer matrix $\tau_\eta$.

\begin{equation}
\centering
\tau_{\eta}=\frac{1}{\sqrt{2}}\left[ \begin{matrix}
   \sqrt{1-\eta} & \sqrt{\eta}  \\
   \sqrt{\eta} & -\sqrt{1-\eta}  \\
\end{matrix} \right],
\end{equation}

where $\eta$ is the probability to the lose a photon in the delay loop. Taking into account the lossy beam splitter we find the following recurrence relations: 

\begin{equation}
\centering
\hat{a}^{\dagger}_{3,j}=\sqrt{1-\eta}\hat{a}^{\dagger}_{0,j+1}+\sqrt{\eta}\hat{a}^{\dagger}_{lost}.
\end{equation}
\begin{equation}
\centering
\hat{a}^{\dagger}_{0,j}=\frac{1}{\sqrt{2}}(\hat{a}^{\dagger}_{2,j}+i\sqrt{1-\eta}\hat{a}^{\dagger}_{0,j+1}+i\sqrt{\eta}\hat{a}^{\dagger}_{lost}).
\end{equation}
For the operator going to the detector (denoted as $\hat{a}^{\dagger}_{2}$ in Fig.~\ref{compare} (b)), we keep calculating the initial input operator $\hat{a}^{\dagger}_{1,j}$ denoted by $\hat{a}^{\dagger}_{2}$ in different time-bin $n$. Then we have:

\begin{equation}
\small
\centering
\begin{split}
\hat{a}^{\dagger}_{1,j}=&\sum_{n\geqslant1}\left( \frac{i^{n-1}(1-\eta)^{\frac{n}{2}}}{2^{\frac{n+1}{2}}}\hat{a}^{\dagger}_{2,j+n}+\frac{i^n \eta^{\frac{1}{2}}(1-\eta)^{\frac{n}{2}}}{2^{\frac{n+1}{2}}}\hat{a}^{\dagger}_{lost}   \right)+\\
&\frac{i\hat{a}^{\dagger}_{2,j}}{\sqrt{2}} +\frac{\sqrt{\eta}}{\sqrt{2}}\hat{a}^{\dagger}_{lost}.
\end{split}
\end{equation}

Since $\hat{a}^{\dagger}_{1,j}$ is carrying the qubit state $S_j$, one can calculate the final state as:

\begin{widetext}
\begin{equation}
\centering
\begin{split}
\ket{\psi}= &(\frac{i}{2^{\frac{1}{2}}}\alpha_0 S_0)\ket{0}_{ph} + (\frac{i}{2^{\frac{1}{2}}}\alpha_1 S_1 + \frac{(1-\eta)^{\frac{1}{2}}}{2}\alpha_0 S_0)\ket{1}_{ph} + (\frac{i}{2^{\frac{1}{2}}}\alpha_2 S_2 + \frac{(1-\eta)^{\frac{1}{2}}}{2}\alpha_1 S_1 + \frac{i(1-\eta)}{2^{\frac{3}{2}}}\alpha_0 S_0)\ket{2}_{ph} + \\
            &(\frac{i}{2^{\frac{1}{2}}}\alpha_3 S_3 + \frac{(1-\eta)^{\frac{1}{2}}}{2}\alpha_2 S_2 + \frac{i(1-\eta)}{2^{\frac{3}{2}}}\alpha_1 S_1-\frac{(1-\eta)^{\frac{3}{2}}}{4}\alpha_0 S_0)\ket{3}_{ph}  + \cdots + \\
            & (\frac{i}{\sqrt{2}}\alpha_{x-1}S_{x-1}+\sum_{n=1}^{x-1}\frac{i^{n-1}(1-\eta)^{\frac{n}{2}}}{2^{\frac{n+1}{2}}}\alpha_{x-n-1}S_{x-n-1})\ket{x-1}_{ph} + (\sum_{n=1}^{x}\frac{i^{n-1}(1-\eta)^{\frac{n}{2}}}{2^{\frac{n+1}{2}}}\alpha_{x-n}S_{x-n})\ket{x}_{ph} + \\
            &(\sum_{n=1}^{x}\frac{i^{n}(1-\eta)^{\frac{n+1}{2}}}{2^{\frac{n+2}{2}}}\alpha_{x-n}S_{x-n})\ket{x+1}_{ph} + \cdots + \\
            &(\sum_{n=1}^{x}\frac{i^{u+n-x-1}(1-\eta)^{\frac{u+n-x}{2}}}{2^{\frac{u+n-x+1}{2}}}\alpha_{x-n}S_{x-n})\ket{u}_{ph} + \cdots + \text{(loss terms)} \\
         =& (\text{terms before $\ket{x}_{ph}$}) + \sum_{u = x}^{\infty}\frac{i^{u-x}(1-\eta)^{\frac{u-x}{2}}}{2^{\frac{u-x}{2}}}(\sum_{n=1}^{x}\frac{i^{n-1}(1-\eta)^{\frac{n}{2}}}{2^{\frac{n+1}{2}}}\alpha_{x-n} S_{x-n})\ket{u}_{ph}  +\text{(loss terms)},
\label{calcs}
\end{split}
\end{equation}
\end{widetext}

where we denote $x=2^m$. Obviously, only when the detector clicks at or after time-bin $x$, the final qubit states can be written in a nicely formed summation without information loss. In order to get the compensation $\alpha_{n}$, we first define:

\begin{equation}
\centering
H(n) = \frac{i^{n-1}(1-\eta)^{\frac{n}{2}}}{2^{\frac{n+1}{2}}},
\end{equation}

and it must satisfy the following relation:

\begin{equation}
\centering
\begin{split}
&|H(1)|^2\alpha_{x-1}^2 = |H(2)|^2\alpha_{x-2}^2 = |H(3)|^2\alpha_{x-3}^2 \\
&= \cdots = |H(x)|^2\alpha_{0}^2.
\end{split}
\end{equation}

Therefore, we have:

\begin{equation}
\centering
\alpha_{n}^2 = \frac{(\frac{2}{1-\eta})^{x-n-1}}{\sum_{j=0}^{x-1}(\frac{2}{1-\eta})^{j}}.
\end{equation}

In this way, we can successfully compensate the losses. The extra phases can be corrected by applying phase gates on Bob's qubits. Therefore, we finally achieve $m$ copies of the Bell state $\ket{\Phi^+}=(\ket{0_A0_B}+\ket{1_A1_B})/\sqrt{2}$. Note that this photon erasure approach is probabilistic instead of deterministic, unlike the one introduced in \cite{zheng2022entanglement}. According to \ref{calcs}, the success probability $P_{succ}$ for the switch-free implementation is:

\begin{equation}
\centering
\begin{split}
P_{succ}(m) &= \sum_{k = 0}^{\infty}\frac{(1-\eta)^{k}}{2^{k}}\left(\sum_{n=1}^{2^m}\frac{(1-\eta)^{n}}{2^{n+1}}\alpha_{2^m-n}^2\right)\\
&= \sum_{k=0}^{\infty}\frac{2^{m-k-2}(1-\eta)^{k+1}}{\sum_{j=0}^{2^m-1}\left(\frac{2}{1-\eta}\right)^{j}}.
\end{split}
\end{equation}

\subsubsection{Concatenated Delay Loops}

As it is shown in the main text, the $P_{succ}$ in the single loop photon erasure is too low. Therefore, one may connect multiple loops and apply the detection at the end, as it is shown in the bottom of Fig.~\ref{compare}. Suppose there are $s$ photon loops connected, and we denote the operators in each loop as $\widehat{a^{(0)}}^{\dagger}$, $\widehat{a^{(1)}}^{\dagger}$, $\widehat{a^{(2)}}^{\dagger}$,\dots, $\widehat{a^{(s-1)}}^{\dagger}$. Note that we have:

\begin{equation}
\centering
\widehat{a^{(k)}}_{2,j}^{\dagger} = \widehat{a^{(k+1)}}_{1,j}^{\dagger},
\label{connector}
\end{equation}

where $k\in\{0,1,2,3,\dots,s-2\}$. By combining the calculations for the single loop and using \ref{connector}, we are able to deduce that the input $\widehat{a^{(0)}}^{\dagger}_{1,j}$ satisfies the Pascal's triangle pattern:

\begin{widetext}
When $s=1$, the detection happens on $\widehat{a^{(0)}}^{\dagger}_{2}$:

\begin{equation}
\widehat{a^{(0)}}^{\dagger}_{1,j}=\frac{i}{\sqrt{2}}\widehat{a^{(0)}}^{\dagger}_{2,j}+\sum_{n^{(0)}\geqslant1}H(n^{(0)})\widehat{a^{(0)}}^{\dagger}_{2,j+n^{(0)}} + \text{loss terms}.
\end{equation}

When $s=2$, the detection happens on $\widehat{a^{(1)}}^{\dagger}_{2}$:
\begin{equation}
\begin{split}
\widehat{a^{(0)}}^{\dagger}_{1,j}&=\frac{i}{\sqrt{2}}\widehat{a^{(0)}}^{\dagger}_{2,j}+\sum_{n^{(0)}\geqslant1} H(n^{(0)})\widehat{a^{(0)}}^{\dagger}_{2,j+n^{(0)}}   + \text{loss terms} \\
&=\frac{i}{\sqrt{2}}\widehat{a^{(1)}}^{\dagger}_{1,j}+\sum_{n^{(0)}\geqslant1} H(n^{(0)})\widehat{a^{(1)}}^{\dagger}_{1,j+n^{(0)}}   + \text{loss terms} \\
&=\left(\frac{i}{\sqrt{2}}\right)^{2}\widehat{a^{(1)}}^{\dagger}_{2,j}+2\frac{i}{\sqrt{2}}\sum_{n^{(0)}\geqslant1}H(n^{(0)})\widehat{a^{(1)}}^{\dagger}_{2,j+n^{(0)}}+\sum_{n^{(0)},n^{(1)}\geqslant1}H(n^{(0)})H(n^{(1)})\widehat{a^{(1)}}^{\dagger}_{2,j+n^{(0)}+n^{(1)}} + \text{loss terms}.
\end{split}
\end{equation}

When $s=3$, the detection happens on $\widehat{a^{(2)}}^{\dagger}_{2}$:

\begin{equation}
\begin{split}
\widehat{a^{(0)}}^{\dagger}_{1,j}=&\left(\frac{i}{\sqrt{2}}\right)^{2}\widehat{a^{(1)}}^{\dagger}_{2,j}+2\frac{i}{\sqrt{2}}\sum_{n^{(0)}\geqslant1}H(n^{(0)})\widehat{a^{(1)}}^{\dagger}_{2,j+n^{(0)}}+\sum_{n^{(0)},n^{(1)}\geqslant1}H(n^{(0)})H(n^{(1)})\widehat{a^{(1)}}^{\dagger}_{2,j+n^{(0)}+n^{(1)}} + \text{loss terms} \\
=&\left(\frac{i}{\sqrt{2}}\right)^{2}\widehat{a^{(2)}}^{\dagger}_{1,j}+2\frac{i}{\sqrt{2}}\sum_{n^{(0)}\geqslant1}H(n^{(0)})\widehat{a^{(2)}}^{\dagger}_{1,j+n^{(0)}}+\sum_{n^{(0)},n^{(1)}\geqslant1}H(n^{(0)})H(n^{(1)})\widehat{a^{(2)}}^{\dagger}_{1,j+n^{(0)}+n^{(1)}} + \text{loss terms} \\
=&\left(\frac{i}{\sqrt{2}}\right)^3\widehat{a^{(2)}}^{\dagger}_{2,j}+3\left(\frac{i}{\sqrt{2}} \right)^2\sum_{n^{(0)}\geqslant1}H(n^{(0)})\widehat{a^{(2)}}^{\dagger}_{2,j+n^{(0)}}+3\frac{i}{\sqrt{2}}\sum_{n^{(0)},n^{(1)}\geqslant1}H(n^{(0)})H(n^{(1)})\widehat{a^{(2)}}^{\dagger}_{2,j+n^{(0)}+n^{(1)}} \\
&+\sum_{n^{(0)},n^{(1)},n^{(2)}\geqslant1}H(n^{(0)})H(n^{(1)})H(n^{(2)})\widehat{a^{(2)}}^{\dagger}_{2,j+n^{(0)}+n^{(1)}+n^{(2)}}+\text{loss terms}.
\end{split}
\end{equation}

One can continue the calculations and get the general formula for any $s\in\mathbb{Z^+}$:

\begin{equation}
\centering
\begin{split}
&\widehat{a^{(0)}}^{\dagger}_{1,j}=\binom{s}{0}\left( \frac{i}{\sqrt{2}}\right)^{s}\widehat{a^{(s-1)}}^{\dagger}_{2,j}+ \sum_{d=1}^{s}\binom{s}{d}\left( \frac{i}{\sqrt{2}} \right)^{s-d}\left[\sum_{n^{(0)},\dots,n^{(d-1)}=1}^{\infty}\left(  \prod_{\tau=0}^{d-1}H(n^{(\tau)}) \right)\widehat{a^{(s-1)}}^{\dagger}_{2,j+\sum_{k=0}^{d-1}n^{(k)}}\right].
\end{split}
\label{beforeaction}
\end{equation}

Now, we perform the actions on the state, i.e., the final state, after the detector clicks. According to the state before photon measurement in \ref{nnnnn}, we can calculate the resulting state for each clicking moment. Similar to the single loop scenario, we obtain the final state:

\begin{equation}
\centering
\begin{split}
\ket{\psi}=&\text{(terms before $\ket{x}_{ph}$)}+\left(\frac{i}{\sqrt{2}}\right)^{s-1}\sum_{u=x}^{\infty} \left\{ \sum_{n=0}^{x-1}\left[   \sum_{k=1}^{\min[s,u-n]}(-1)^{k+1}\binom{s}{k}\binom{u-n-1}{k-1}     \right] \right.  \left. H(u-n)\alpha_{n}S_{n}\right\}\ket{u}_{ph} \\
&+\text{(loss terms)} \\ 
=&\text{(terms before $\ket{x}_{ph}$)}+ \left(\frac{i}{\sqrt{2}}\right)^{s-1}\sum_{u=x}^{\infty}\left\{\sum_{n=0}^{x-1}\left[ \boldsymbol{Y}(s,u,n)H(u-n)\alpha_{n}S_{n} \right]\right\}\ket{u}_{ph} +\text{(loss terms)} \\
=&\text{(terms before $\ket{x}_{ph}$)}+\sum_{u=x}^{\infty}\left\{ \left(\frac{i^{s+u-2}\sqrt{1-\eta}^u}{\sqrt{2}^{s+u}}\right)\sum_{n=0}^{x-1}\left(\frac{\sqrt{2}}{i\sqrt{1-\eta}}\right)^n\boldsymbol{Y}(s,u,n)\alpha_nS_n \right\}\ket{u}_{ph} +\text{(loss terms)},
\end{split}
\label{connectaction}
\end{equation}

where we define:

\begin{equation}
\boldsymbol{Y}(s,u,k)=\sum_{t=1}^{\min[s,u-k]}(-1)^{t+1}\binom{s}{t}\binom{u-k-1}{t-1}.
\end{equation}

\end{widetext}

Note that the index $k$ is upper bounded by $\min[s,u-n]$ since practically, we do not have an infinitely large number of loops $s$. Thus, some of the terms in \eqref{beforeaction} may be truncated. Still, we have the terms in a nicely-formed summation until the clicking time-bin is at or after $\ket{x}_{ph}$. However, it does not guarantee that we have good states. In order to make the state without information loss and phase correctable (i.e., the terms before each $S_n$ are non-zero and have the same sign), we must choose a suitable integer set $(s,u)$ such that $\boldsymbol{Y}(s,u,n)$ has the same sign throughout all $n$, i.e.,

\begin{equation}
\begin{split}
&\boldsymbol{Y}(s,u,n)>0 \ \text{or} \ \boldsymbol{Y}(s,u,n)<0  \\
&\text{for} \ \forall n\in\{0,1,2,\dots,x-1\}.
\end{split}
\end{equation}

In this way, we can compensate the amplitude of the photonic pulses, which should satisfy:

\begin{equation}
\begin{split}
&|\boldsymbol{Y}(s,u,0) \frac{(1-\eta)^{-\frac{0}{2}}}{2^{-\frac{0}{2}}}\alpha_0|^2 = |\boldsymbol{Y}(s,u,1) \frac{(1-\eta)^{-\frac{1}{2}}}{2^{-\frac{1}{2}}}\alpha_1|^2 = \\
&|\boldsymbol{Y}(s,u,2) \frac{(1-\eta)^{-\frac{2}{2}}}{2^{-\frac{2}{2}}}\alpha_2|^2=\cdots=\\
&|\boldsymbol{Y}(s,u,x-1) \frac{(1-\eta)^{-\frac{x-1}{2}}}{2^{-\frac{x-1}{2}}}\alpha_{x-1}|^2.
\end{split}
\end{equation}

Therefore, according to the equation above and the normalization condition $\sum_{n=0}^{2^m-1}\alpha_n^2=1$, we can obtain the amplitude compensation term:

\begin{equation}
\alpha_{n}^2=\frac{1}{\sum_{k=0}^{x-1}\left|\frac{\boldsymbol{Y}(s,u,n)}{\boldsymbol{Y}(s,u,k)}\right|^2\left(\frac{1-\eta}{2}\right)^{k-n}}.
\label{compen-loss}
\end{equation}

In this way, when the detector clicks at time-bin $u$, we can acquire the state \eqref{ABphase} and transform it to $m$ copies of the Bell state by applying phase correction single-qubit gates. According to \eqref{connectaction}, when we fix the number of connected loops $s$ and the clicking time-bin $u$, the corresponding photon erasure success probability (regardless of qubit-qudit entanglement local losses and transmission loss) is:

\begin{equation}
P_{succ}(s,u,m) = \frac{x(1-\eta)^u}{2^{s+u}}\frac{1}{\sum_{k=0}^{x-1}\left|\frac{1}{\boldsymbol{Y}(s,u,k)}\right|^2\left(\frac{1-\eta}{2}\right)^k}.
\label{succ-mult}
\end{equation}

Note that in this case, since $\alpha_n$ depends on $u$ and $s$ and we need to generate the pulses according to $\alpha_n$ at the very beginning of the protocol, we need to choose only one clicking time-bin as the success flag. Therefore, one needs to find the best $(s,u)$ set that outputs the maximum $P_{succ}(s,u,m)$. %We also illustrate an example to see which $(s,u)$ sets can output the maximum $P_{succ}(s,u,m)$, as shown in Table~\ref{tableprob}.

% \begin{table}[]
% \centering
% \begin{tabular}{|c|c|c|}
% \hline
%  & \begin{tabular}[c]{@{}c@{}}Maximum Success \\ probability for $s\geqslant2$\end{tabular} & \begin{tabular}[c]{@{}c@{}}Success Probability \\ for $s=1$ $(u\geqslant2^m)$\end{tabular} \\ \hline
% m=2 & $2.62\times10^{-2}$ $(s=2, u=8)$  & $1.22\times10^{-1}$ \\ \hline
% m=3 & $1.45\times10^{-2}$ $(s=2, u=11)$ & $1.33\times10^{-2}$ \\ \hline
% m=4 & $2.08\times10^{-3}$ $(s=5,u=32)$  & $8.84\times10^{-5}$ \\ \hline
% m=5 & $6.63\times10^{-5}$ $(s=11,u=78)$ & $1.95\times10^{-9}$ \\ \hline
% \end{tabular}
% \caption{A comparison of the photon erasure success probability $P_{succ}$ for the single loop and the concatenated loops cases. We suppose $\eta=2\%$, $s=2\sim15$ and $u=x\sim x+50$. Note that higher values of the success probability for the connected loops may exist outside the selected range for $(s,u)$.}
% \label{tableprob}
% \end{table}

Additionally, the above calculations for the concatenated loops also fit the case for the single loop where $s=1$. The only change to make is that one needs to sum over all the success probabilities $P_{succ}$ for the clicking time $u\geqslant x$ as the erasure success probability for the single loop, i.e.,

\begin{equation}
P_{succ}(s=1, m) = \sum_{u=x}^{\infty}P_{succ}(s=1,u,m). 
\label{succ-1}
\end{equation}

\subsubsection{Photon Measurement with Loss}

Generally, the loss of the photon can happen during the local qubit-qudit interaction, the transmission between Alice and Bob, and the delay loop of the photon erasure. 

The fidelity of the resulting state will not be affected by the previous two losses since the photon is subject to the same amount of losses regardless of which time-bin it is in.  Therefore, they do not contribute any correlated errors but will only lower the success probability of the protocol. The local losses increases with the number of entangled pairs such that the total loss probability from the qudit-qubit interactions is $(1-\eta_{AB})^m$, where $\eta_{AB}$ is the loss per interaction. The transmission probability between Alice and Bob is denoted as $1-\eta_{0}$ and we assume fiber loss at telecom frequencies.

\begin{figure}
    \includegraphics[width=1.0\linewidth]{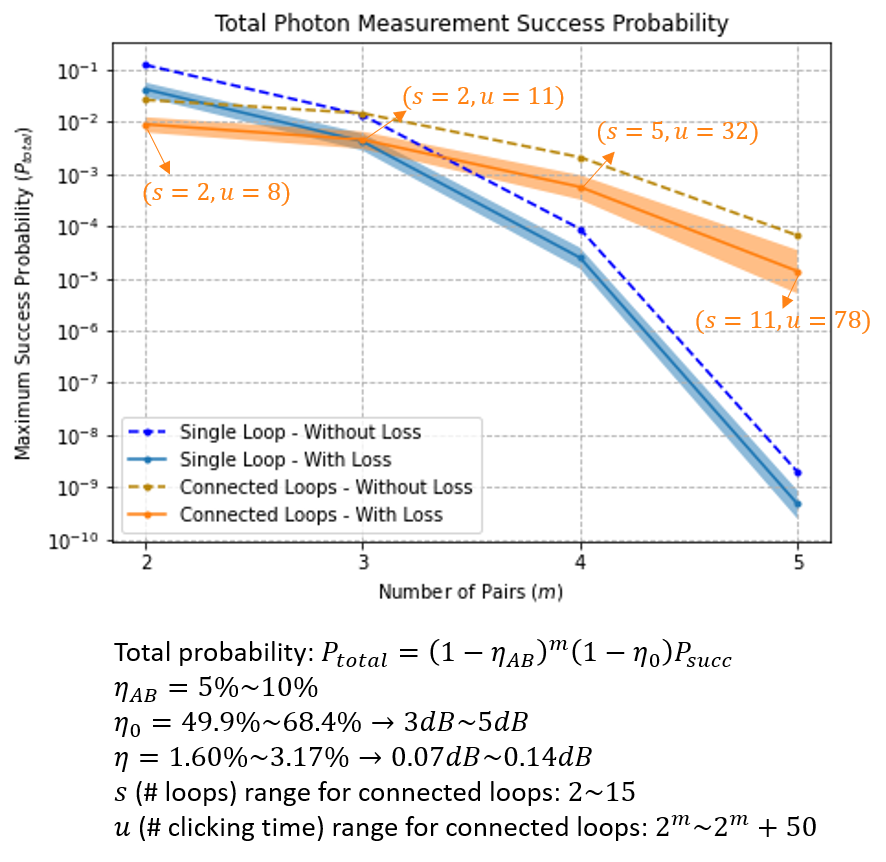}
    \caption{The total success probability of the photon erasure with respect to the number of generated pairs ($m=2\sim5$) is shown. We consider both the single loop and the connected loops cases. We define the possible parameter ranges for the losses, $s$ and $u$, and then find the maximum $P_{total}$. The applied $(s,u)$-s are marked on the plot. Connecting photon loops increases the total measurement success probability compared with the case where only one single photon loop is placed. Note that for single photon loop, the success probability refers to \eqref{succ-1}. The cases where no loss occurs are also plotted for the comparisons.}  
    \label{succprob}  
\end{figure}

For the third kind of photon losses - the losses in the delay loops of the photon erasure, they lower the success probability but they also affect the fidelity of the generated pairs since each qudit pulse may also experience a different number of delay loops. Nevertheless, we can choose the compensation coefficients $\alpha_n$ in Eqs. \ref{compen-loss}, which will vary with the loss on each delay loop $\eta$ as well. As a result, the success probability for photon measurement will change correspondingly. The overall success probability $P_{total}$ is:

\begin{equation}
\centering
P_{total}=(1-\eta_{AB})^m(1-\eta_{0})P_{succ},
\end{equation}

where $P_{succ}$ is defined in \eqref{succ-1} for $s=1$ and \eqref{succ-mult} for $s\geq2$. Now, we can input some moderate loss parameters for $\eta$, $\eta_{AB}$ and $\eta_0$ to discuss the $P_{total}$.

Firstly, we assume the photon wavelength is in the range of $600\sim800$nm corresponding to typical optical transitions of realevant quantum hardware such as group-IV diamond defect centers. For the transmission betweeen Alice and Bob, the photon will have to be frequency converted to the telecom band for efficient transmission in low loss optical fibers before being converted back to $600\sim800$nm for interaction with Bob's qubits. We assume that after interaction with Bob's qubits, the wavelength is kept at $600\sim800$nm for the erasure measurement. Assuming fiber delay lines, the loss is $5\sim10$dB/km~\cite{dbloss} for this wavelength range. 

Assuming that the time-bin duration is approximately $70$ns~\cite{bhaskar2020experimental, knall2022efficient} and the transmission velocity of the photon in the fiber is $2\times10^5$km/s, we find that the dB loss on each delay loop is around $0.07\sim0.14$dB and the corresponding loss percentage is $\eta=0.0160\sim0.0317$.

Assuming frequency conversion to the telecom band for the transmission between Alice and Bob the transmission loss will be  $0.15\sim0.25$dB/km \cite{li2020advances} and resulting in a transmission of $\eta_{0}=0.499\sim0.684$ for 20km separation. We also assume $\eta_{AB}=0.05\sim0.10$ for a general approximation of the local losses.

With these parameters and according to $P_{total}$, one can find the total photon measurement success probability as shown in Fig.~\ref{succprob}. We can find that $P_{total}$ will, of course, drop when the number of generated pairs increases. Moreover, for the single photon loop, the $P_{total}$ drops drastically, which will make the measurement vulnerable to the detector's dark count rate (around $10^{-7}\sim10^{-6}$). However, with photon loops connected, the drop of the $P_{total}$ can be slightly mitigated, making the measurement more resistant to the dark count.

\subsection{Quantum Memory}

Qubits are vulnerable to decoherence. To simulate qubit decoherence during transmission, we employ a dephasing and generalized amplitude damping channel. Suppose the distance between Alice and Bob is $L$ the speed of photon transmission between them is represented by $c$.  In this scenario, Alice must send the photon to Bob and wait for the heralding information to return.  Additionally, we define the success of the protocol based on Alice receiving Bob's heralding message. Therefore, the waiting time $t$ for Alice is:

\begin{equation}
\centering
t_A=\frac{2L}{c},
\end{equation}

and for Bob:

\begin{equation}
\centering
t_B=\frac{L}{c}.
\end{equation}

Suppose the dephasing time is $T_p$, then the Kraus operators of the dephasing channel are:

\begin{equation}
\centering
\begin{split}
&A_{0}=\sqrt{\frac{1+e^{-t/T_{p}}}{2}}\boldsymbol{I},\\
&A_{1}=\sqrt{\frac{1-e^{-t/T_{p}}}{2}}\boldsymbol{Z},
\end{split}
\end{equation}

where $\boldsymbol{I}$ and $\boldsymbol{Z}$ denote the Identity and Pauli-Z matrix respectively. Also, by indicating the relaxation time $T_1$, the Kraus operators of the amplitude damping channel are:

\begin{equation}
\centering
\begin{split}
&{{E}_{0}}=\sqrt{{{a}_{\beta }}}\left[ \begin{matrix}
   1 & 0  \\
   0 & {{e}^{-t/2{{T}_{1}}}}  \\
\end{matrix} \right], \\
&{{E}_{1}}=\sqrt{{{a}_{\beta }}}\left[ \begin{matrix}
   0 & \sqrt{1-{{e}^{-t/{{T}_{1}}}}}  \\
   0 & 0  \\
\end{matrix} \right], \\
&{{E}_{2}}=\sqrt{1-{{a}_{\beta }}}\left[ \begin{matrix}
   {{e}^{-t/2{{T}_{1}}}} & 0  \\
   0 & 1  \\
\end{matrix} \right], \\
&{{E}_{3}}=\sqrt{1-{{a}_{\beta }}}\left[ \begin{matrix}
   0 & 0  \\
   \sqrt{1-{{e}^{-t/{{T}_{1}}}}} & 0  \\
\end{matrix} \right], \\
\end{split}
\end{equation}

where:

\begin{equation}
\centering
a_{\beta}=e^{-\beta\Delta E}=e^{-\frac{\Delta E}{k_B T}},
\end{equation}

and $\Delta E$, $k_B$ and $T$ denote the energy difference between $\ket{0}$ and $\ket{1}$, Boltzmann constant and system temperature, respectively. Therefore, by applying the two kinds of Kraus operators, the output density matrix $\rho$ becomes:

\begin{equation}
\centering
\rho'=\sum_{i=0,1;j=0,1,2,3}A_i E_j \rho E_j^{\dagger} A_i^{\dagger}.
\label{channel}
\end{equation}

We set $L=20km$, $T_1=10ms$, $T_p=5ms$ and $a_{\beta}=0.5$. Note that this is the case for only one qubit. For a multi-qubit system, since the decoherence channels on each qubit are uncorrelated, we can apply Eqs. (\ref{channel}) on each qubit independently.

Notably, despite the errors mentioned above, qubit gate errors should also be considered, especially for the latter one - since we put a lot of Hadamard operations inside.

\section{\label{B}Genetic Purification Protocol}

\begin{figure*}
    \centering
    \includegraphics[width=0.85\linewidth]{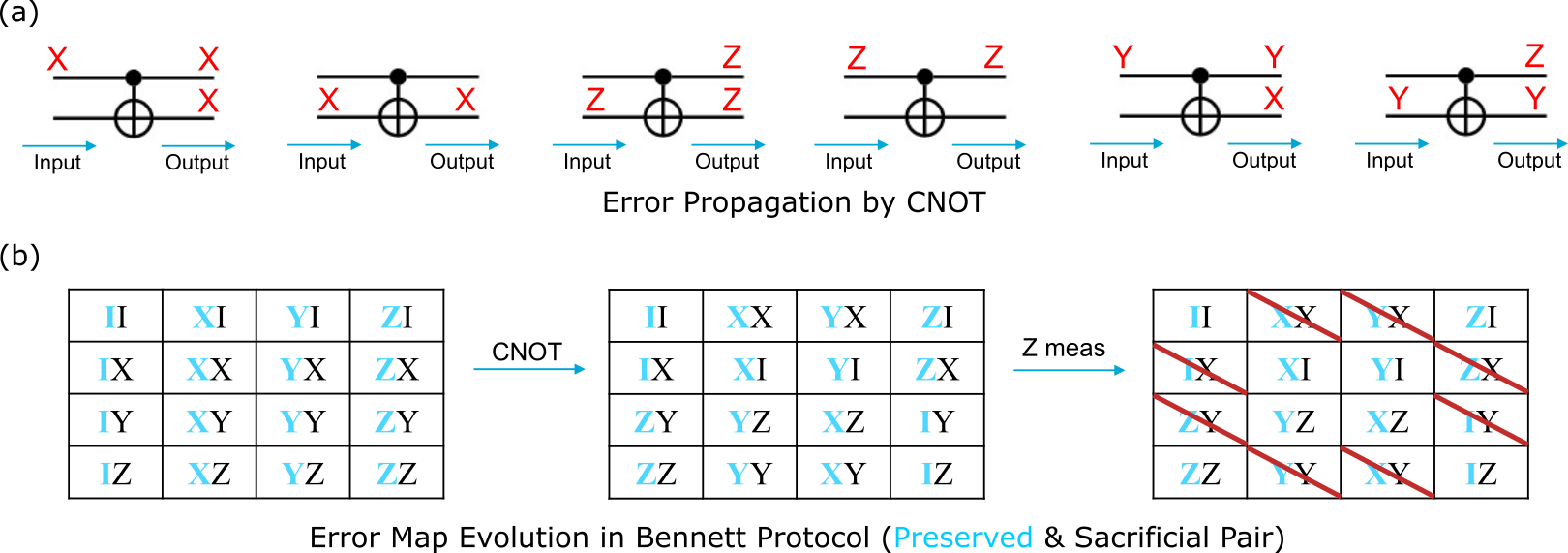}
    \caption{(a): how CNOT propagates the input errors. (b): The error map evolution within Bennett protocol \cite{bennett1996purification}.} 
    \label{prop}
\end{figure*}

In this appendix section, we show the basic idea of the optimized purification protocol proposed by \cite{krastanov2019optimized}.

\subsection{Error Detection by Measurement}

Consider the noises on the entangled pairs are described as depolarizing errors:

\begin{equation}
\centering
\begin{split}
\rho_{AB}=&(1-x_0-y_0-z_0)\ket{\Phi^+}\bra{\Phi^+}+x_0\ket{\Psi^+}\bra{\Psi^+} \\
&+y_0\ket{\Psi^-}\bra{\Psi^-}+z_0\ket{\Phi^-}\bra{\Phi^-}.
\label{B1}
\end{split}
\end{equation}

Similar to the conventional depolarizing channel, we input $I$ (no errors), $X$, $Y$, or $Z$ errors into the perfect state with the corresponding possibilities. For the Bell pair case, the desired perfect state is $\ket{\Phi^+}$ and the other three erroneous states are:

\begin{equation}
\label{errorbasis}
\centering
\begin{split}
&\ket{\Psi^+}=\frac{1}{\sqrt{2}}(\ket{01}+\ket{10})\rightarrow \boldsymbol{X} \ \text{error},\\
&\ket{\Psi^-}=\frac{1}{\sqrt{2}}(\ket{01}-\ket{10})\rightarrow \boldsymbol{Y} \ \text{error},\\
&\ket{\Phi^-}=\frac{1}{\sqrt{2}}(\ket{00}-\ket{11})\rightarrow \boldsymbol{Z} \ \text{error}.
\end{split}
\end{equation}

Note that in the output states, there may also exist some coherent errors (e.g., $\ket{\Phi^+}\bra{\Psi^-}$). These errors can be eliminated by using the twirling error mitigation technique \cite{geller2013efficient}. However, for the discussions in our work, we will not add the twirling process but implement the purification only.

If Alice and Bob then measure their own register qubits separately, they will know if their previous shared entangled pair has errors or not. There are three possible measurement blocks that Alice and Bob can choose:

\begin{itemize}
\setlength{\itemsep}{0pt}
\item Measurement in $\boldsymbol{X}$ basis,
\item Measurement in $\boldsymbol{Y}$ basis,
\item Measurement in $\boldsymbol{Z}$ basis,
\end{itemize}

and there are two possible measurement results from Alice and Bob:

\begin{itemize}
\setlength{\itemsep}{0pt}
\item Coincidence Result: $0_A0_B$ and $1_A1_B$,
\item Anti-coincidence Result: $0_A1_B$ and $1_A0_B$.
\end{itemize}

Note that the measurement on one certain basis can only detect two kinds of errors. For example, by applying $\boldsymbol{X}$-basis measurement, only $\boldsymbol{Y}$ and $\boldsymbol{Z}$ can be detected. Also, one should always preserve $\ket{\Phi^+}$ and cannot consider it as one of the erroneous states. Therefore, if Alice and Bob want to know whether their previous shared $\rho_{AB}$ is possibly in $\ket{\Phi^+}$, their measurement results and measurement basis should obey the criteria as follows:

\begin{itemize}
\setlength{\itemsep}{0pt}
\item Measurement in $\boldsymbol{X}$ basis: Coincidence result $\rightarrow$ $\ket{\Psi^-}$ and $\ket{\Phi^-}$ detectable.
\item Measurement in $\boldsymbol{Y}$ basis: Anti-coincidence result $\rightarrow$ $\ket{\Psi^+}$ and $\ket{\Phi^-}$ detectable.
\item Measurement in $\boldsymbol{Z}$ basis: Coincidence result $\rightarrow$ $\ket{\Psi^-}$ and $\ket{\Psi^+}$ detectable.
\end{itemize}

If the measurement result of Alice and Bob violates the criteria above, Alice and Bob will know that their previous shared pair $\rho_{AB}$ is in one of the erroneous states.

\subsection{Error Propagation}

By directly measuring the entangled pair, we are indeed able to detect the errors inside the Bell pair, but in the meantime, also destroy the entanglement. In order to make errors detectable without entanglement demolition, one should introduce another sacrificial entangled pair. Alice and Bob can then apply CNOTs locally to connect the preserved and sacrificial pair and propagate the errors in the preserved pair to the sacrificial pair as shown in Fig.~\ref{prop} (a). The final measurement check will be applied to the sacrificial pairs to detect and eliminate errors.

Consider the simplest purification protocol, the Bennett protocol \cite{bennett1996purification}. Suppose both two pairs are modeled as the depolarizing channel as shown in Eqs. (\ref{B1}), and the error map is shown in Fig.~\ref{prop} (b). The CNOT will change the error map according to the error propagation in Fig.~\ref{prop} (a). The final Z-basis measurement will eliminate the blocks with X and Y errors on the sacrificial pair.

\subsection{Error Permutation}

The error propagation indeed provides the approach to send the larger errors to the sacrificial pair and eliminate them by measurements. However, the freedom of controlling the error flow is still limited. 

\begin{figure*}
    \centering
    \includegraphics[width=0.95\linewidth]{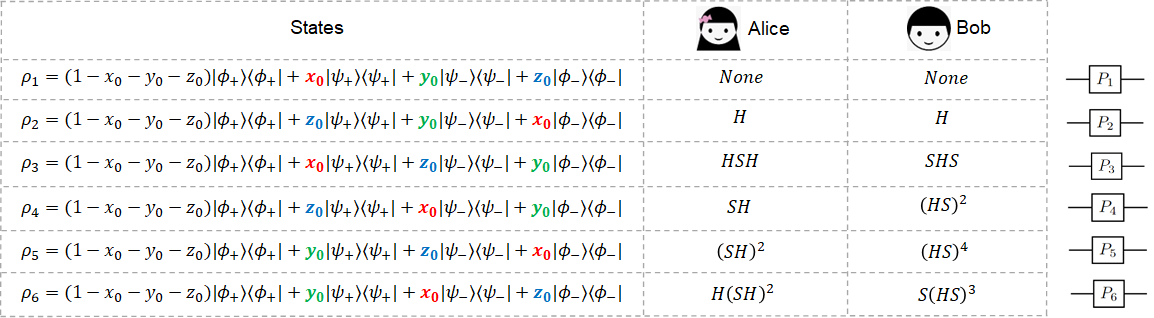}
    \caption{The error permutation and the corresponding single-qubit gate operations applied on Alice and Bob.} 
    \label{permu}
\end{figure*}

The error permutation is a technique that permutes the error coefficients by letting Alice and Bob apply corresponding operations locally. The coefficient permutation and the corresponding single-qubit gate operations on Alice and Bob are shown in Fig.~\ref{permu}. For example, if Alice and Bob apply nothing, the density matrix will remain the same as Eqs. \eqref{B1}. However, if Alice and Bob apply a Hadamard gate respectively, the shared density matrix will become:

\begin{equation}
\centering
\begin{split}
\rho_{AB}=&(1-x_0-y_0-z_0)\ket{\Phi^+}\bra{\Phi^+}+z_0\ket{\Psi^+}\bra{\Psi^+}\\
&+y_0\ket{\Psi^-}\bra{\Psi^-}+x_0\ket{\Phi^-}\bra{\Phi^-},
\end{split}
\end{equation}

which applies the exchange between $x_0$ and $z_0$. Suppose $z_0$ is much larger than $x_0$ and $z_0$ and we still keep the Z-basis measurement. In Eqs. \eqref{B1}, we can only eliminate $\ket{\Psi^+}$ and $\ket{\Psi^-}$ affiliated with $x_0$ and $y_0$. However, if we apply the permutation shown above, still, $\ket{\Psi^+}$ and $\ket{\Psi^-}$ will be eliminated but they are affiliated with $z_0$ and $y_0$. In this way, we flow the erroneous states with the largest coefficient into the error detection part without changing the measurement basis.

\subsection{Evolutionary Algorithm}

We have different measurement blocks, error propagation, and permutation techniques already. Now, we should assemble these components to find the optimal purification circuit. The evolutionary algorithm is adopted to search for an optimized purification protocol. The procedure is shown as follows:

\begin{itemize}
\setlength{\itemsep}{0pt}
\item Completely randomize the initial circuit group by stochastically assembling the CNOTs, permutation gates, and measurement blocks. 
\item Input the Bell pair state with the specific error model in the purification circuits, and output the fidelity of the purified Bell pair for all purification circuits. Select the best purification circuits (more than one).
\item Consider that the selected circuits are the parent circuits, and the operations on the circuits are their genes. Let the parent circuits reproduce baby circuits by randomly mixing the genes from their parents.
\item The selected circuits and the baby circuits form the new circuit group.
\item Since no new genes (operations) are involved, the mutations should be introduced, which is, some of the operations are mutated to another random different operation with relatively low probabilities among all circuits in the newly-formed group. 
\item Output the purified fidelity in the circuits from the mutated group. 
\item Loop it back to the parent circuit selection step and continue the loop hundreds of times until the maximum purified fidelity reaches convergence.
\item The purification circuit with the converged fidelity will be the optimized purification circuit for the specific error model.
\end{itemize}

Note that all the purified circuits and the following output results are under the condition of the minimum entanglement requirement. The minimum entanglement requirement means using the minimum number of CNOTs to connect all shared Bell pairs without leaving one pair unoperated. One can indeed set the purified circuits to a larger size and run the algorithm, but with more local two-qubit gate consumptions.


\nocite{1}
% \nocite{*}

\bibliography{apssamp}% Produces the bibliography via BibTeX.

\providecommand{\noopsort}[1]{}\providecommand{\singleletter}[1]{#1}%
\begin{thebibliography}{39}%
\makeatletter
\providecommand \@ifxundefined [1]{%
 \@ifx{#1\undefined}
}%
\providecommand \@ifnum [1]{%
 \ifnum #1\expandafter \@firstoftwo
 \else \expandafter \@secondoftwo
 \fi
}%
\providecommand \@ifx [1]{%
 \ifx #1\expandafter \@firstoftwo
 \else \expandafter \@secondoftwo
 \fi
}%
\providecommand \natexlab [1]{#1}%
\providecommand \enquote  [1]{``#1''}%
\providecommand \bibnamefont  [1]{#1}%
\providecommand \bibfnamefont [1]{#1}%
\providecommand \citenamefont [1]{#1}%
\providecommand \href@noop [0]{\@secondoftwo}%
\providecommand \href [0]{\begingroup \@sanitize@url \@href}%
\providecommand \@href[1]{\@@startlink{#1}\@@href}%
\providecommand \@@href[1]{\endgroup#1\@@endlink}%
\providecommand \@sanitize@url [0]{\catcode `\\12\catcode `\$12\catcode
  `\&12\catcode `\#12\catcode `\^12\catcode `\_12\catcode `\%12\relax}%
\providecommand \@@startlink[1]{}%
\providecommand \@@endlink[0]{}%
\providecommand \url  [0]{\begingroup\@sanitize@url \@url }%
\providecommand \@url [1]{\endgroup\@href {#1}{\urlprefix }}%
\providecommand \urlprefix  [0]{URL }%
\providecommand \Eprint [0]{\href }%
\providecommand \doibase [0]{https://doi.org/}%
\providecommand \selectlanguage [0]{\@gobble}%
\providecommand \bibinfo  [0]{\@secondoftwo}%
\providecommand \bibfield  [0]{\@secondoftwo}%
\providecommand \translation [1]{[#1]}%
\providecommand \BibitemOpen [0]{}%
\providecommand \bibitemStop [0]{}%
\providecommand \bibitemNoStop [0]{.\EOS\space}%
\providecommand \EOS [0]{\spacefactor3000\relax}%
\providecommand \BibitemShut  [1]{\csname bibitem#1\endcsname}%
\let\auto@bib@innerbib\@empty
%</preamble>
\bibitem [{\citenamefont {Pirandola}\ \emph {et~al.}(2020)\citenamefont
  {Pirandola}, \citenamefont {Andersen}, \citenamefont {Banchi}, \citenamefont
  {Berta}, \citenamefont {Bunandar}, \citenamefont {Colbeck}, \citenamefont
  {Englund}, \citenamefont {Gehring}, \citenamefont {Lupo}, \citenamefont
  {Ottaviani} \emph {et~al.}}]{pirandola2020advances}%
  \BibitemOpen
  \bibfield  {author} {\bibinfo {author} {\bibfnamefont {S.}~\bibnamefont
  {Pirandola}}, \bibinfo {author} {\bibfnamefont {U.~L.}\ \bibnamefont
  {Andersen}}, \bibinfo {author} {\bibfnamefont {L.}~\bibnamefont {Banchi}},
  \bibinfo {author} {\bibfnamefont {M.}~\bibnamefont {Berta}}, \bibinfo
  {author} {\bibfnamefont {D.}~\bibnamefont {Bunandar}}, \bibinfo {author}
  {\bibfnamefont {R.}~\bibnamefont {Colbeck}}, \bibinfo {author} {\bibfnamefont
  {D.}~\bibnamefont {Englund}}, \bibinfo {author} {\bibfnamefont
  {T.}~\bibnamefont {Gehring}}, \bibinfo {author} {\bibfnamefont
  {C.}~\bibnamefont {Lupo}}, \bibinfo {author} {\bibfnamefont {C.}~\bibnamefont
  {Ottaviani}}, \emph {et~al.},\ }\bibfield  {title} {\bibinfo {title}
  {Advances in quantum cryptography},\ }\href@noop {} {\bibfield  {journal}
  {\bibinfo  {journal} {Advances in optics and photonics}\ }\textbf {\bibinfo
  {volume} {12}},\ \bibinfo {pages} {1012} (\bibinfo {year}
  {2020})}\BibitemShut {NoStop}%
\bibitem [{\citenamefont {Degen}\ \emph {et~al.}(2017)\citenamefont {Degen},
  \citenamefont {Reinhard},\ and\ \citenamefont
  {Cappellaro}}]{degen2017quantum}%
  \BibitemOpen
  \bibfield  {author} {\bibinfo {author} {\bibfnamefont {C.~L.}\ \bibnamefont
  {Degen}}, \bibinfo {author} {\bibfnamefont {F.}~\bibnamefont {Reinhard}},\
  and\ \bibinfo {author} {\bibfnamefont {P.}~\bibnamefont {Cappellaro}},\
  }\bibfield  {title} {\bibinfo {title} {Quantum sensing},\ }\href@noop {}
  {\bibfield  {journal} {\bibinfo  {journal} {Reviews of modern physics}\
  }\textbf {\bibinfo {volume} {89}},\ \bibinfo {pages} {035002} (\bibinfo
  {year} {2017})}\BibitemShut {NoStop}%
\bibitem [{\citenamefont {Zhang}\ and\ \citenamefont
  {Zhuang}(2021)}]{zhang2021distributed}%
  \BibitemOpen
  \bibfield  {author} {\bibinfo {author} {\bibfnamefont {Z.}~\bibnamefont
  {Zhang}}\ and\ \bibinfo {author} {\bibfnamefont {Q.}~\bibnamefont {Zhuang}},\
  }\bibfield  {title} {\bibinfo {title} {Distributed quantum sensing},\
  }\href@noop {} {\bibfield  {journal} {\bibinfo  {journal} {Quantum Science
  and Technology}\ }\textbf {\bibinfo {volume} {6}},\ \bibinfo {pages} {043001}
  (\bibinfo {year} {2021})}\BibitemShut {NoStop}%
\bibitem [{\citenamefont {Guo}\ \emph {et~al.}(2020)\citenamefont {Guo},
  \citenamefont {Breum}, \citenamefont {Borregaard}, \citenamefont {Izumi},
  \citenamefont {Larsen}, \citenamefont {Gehring}, \citenamefont {Christandl},
  \citenamefont {Neergaard-Nielsen},\ and\ \citenamefont
  {Andersen}}]{guo2020distributed}%
  \BibitemOpen
  \bibfield  {author} {\bibinfo {author} {\bibfnamefont {X.}~\bibnamefont
  {Guo}}, \bibinfo {author} {\bibfnamefont {C.~R.}\ \bibnamefont {Breum}},
  \bibinfo {author} {\bibfnamefont {J.}~\bibnamefont {Borregaard}}, \bibinfo
  {author} {\bibfnamefont {S.}~\bibnamefont {Izumi}}, \bibinfo {author}
  {\bibfnamefont {M.~V.}\ \bibnamefont {Larsen}}, \bibinfo {author}
  {\bibfnamefont {T.}~\bibnamefont {Gehring}}, \bibinfo {author} {\bibfnamefont
  {M.}~\bibnamefont {Christandl}}, \bibinfo {author} {\bibfnamefont {J.~S.}\
  \bibnamefont {Neergaard-Nielsen}},\ and\ \bibinfo {author} {\bibfnamefont
  {U.~L.}\ \bibnamefont {Andersen}},\ }\bibfield  {title} {\bibinfo {title}
  {Distributed quantum sensing in a continuous-variable entangled network},\
  }\href@noop {} {\bibfield  {journal} {\bibinfo  {journal} {Nature Physics}\
  }\textbf {\bibinfo {volume} {16}},\ \bibinfo {pages} {281} (\bibinfo {year}
  {2020})}\BibitemShut {NoStop}%
\bibitem [{\citenamefont {Beals}\ \emph {et~al.}(2013)\citenamefont {Beals},
  \citenamefont {Brierley}, \citenamefont {Gray}, \citenamefont {Harrow},
  \citenamefont {Kutin}, \citenamefont {Linden}, \citenamefont {Shepherd},\
  and\ \citenamefont {Stather}}]{beals2013efficient}%
  \BibitemOpen
  \bibfield  {author} {\bibinfo {author} {\bibfnamefont {R.}~\bibnamefont
  {Beals}}, \bibinfo {author} {\bibfnamefont {S.}~\bibnamefont {Brierley}},
  \bibinfo {author} {\bibfnamefont {O.}~\bibnamefont {Gray}}, \bibinfo {author}
  {\bibfnamefont {A.~W.}\ \bibnamefont {Harrow}}, \bibinfo {author}
  {\bibfnamefont {S.}~\bibnamefont {Kutin}}, \bibinfo {author} {\bibfnamefont
  {N.}~\bibnamefont {Linden}}, \bibinfo {author} {\bibfnamefont
  {D.}~\bibnamefont {Shepherd}},\ and\ \bibinfo {author} {\bibfnamefont
  {M.}~\bibnamefont {Stather}},\ }\bibfield  {title} {\bibinfo {title}
  {Efficient distributed quantum computing},\ }\href@noop {} {\bibfield
  {journal} {\bibinfo  {journal} {Proceedings of the Royal Society A:
  Mathematical, Physical and Engineering Sciences}\ }\textbf {\bibinfo {volume}
  {469}},\ \bibinfo {pages} {20120686} (\bibinfo {year} {2013})}\BibitemShut
  {NoStop}%
\bibitem [{\citenamefont {Cacciapuoti}\ \emph {et~al.}(2019)\citenamefont
  {Cacciapuoti}, \citenamefont {Caleffi}, \citenamefont {Tafuri}, \citenamefont
  {Cataliotti}, \citenamefont {Gherardini},\ and\ \citenamefont
  {Bianchi}}]{cacciapuoti2019quantum}%
  \BibitemOpen
  \bibfield  {author} {\bibinfo {author} {\bibfnamefont {A.~S.}\ \bibnamefont
  {Cacciapuoti}}, \bibinfo {author} {\bibfnamefont {M.}~\bibnamefont
  {Caleffi}}, \bibinfo {author} {\bibfnamefont {F.}~\bibnamefont {Tafuri}},
  \bibinfo {author} {\bibfnamefont {F.~S.}\ \bibnamefont {Cataliotti}},
  \bibinfo {author} {\bibfnamefont {S.}~\bibnamefont {Gherardini}},\ and\
  \bibinfo {author} {\bibfnamefont {G.}~\bibnamefont {Bianchi}},\ }\bibfield
  {title} {\bibinfo {title} {Quantum internet: networking challenges in
  distributed quantum computing},\ }\href@noop {} {\bibfield  {journal}
  {\bibinfo  {journal} {IEEE Network}\ }\textbf {\bibinfo {volume} {34}},\
  \bibinfo {pages} {137} (\bibinfo {year} {2019})}\BibitemShut {NoStop}%
\bibitem [{\citenamefont {Buhrman}\ and\ \citenamefont
  {R{\"o}hrig}(2003)}]{buhrman2003distributed}%
  \BibitemOpen
  \bibfield  {author} {\bibinfo {author} {\bibfnamefont {H.}~\bibnamefont
  {Buhrman}}\ and\ \bibinfo {author} {\bibfnamefont {H.}~\bibnamefont
  {R{\"o}hrig}},\ }\bibfield  {title} {\bibinfo {title} {Distributed quantum
  computing},\ }in\ \href@noop {} {\emph {\bibinfo {booktitle} {International
  Symposium on Mathematical Foundations of Computer Science}}}\ (\bibinfo
  {organization} {Springer},\ \bibinfo {year} {2003})\ pp.\ \bibinfo {pages}
  {1--20}\BibitemShut {NoStop}%
\bibitem [{\citenamefont {Awschalom}\ \emph {et~al.}(2021)\citenamefont
  {Awschalom}, \citenamefont {Berggren}, \citenamefont {Bernien}, \citenamefont
  {Bhave}, \citenamefont {Carr}, \citenamefont {Davids}, \citenamefont
  {Economou}, \citenamefont {Englund}, \citenamefont {Faraon}, \citenamefont
  {Fejer}, \citenamefont {Guha}, \citenamefont {Gustafsson}, \citenamefont
  {Hu}, \citenamefont {Jiang}, \citenamefont {Kim}, \citenamefont {Korzh},
  \citenamefont {Kumar}, \citenamefont {Kwiat}, \citenamefont
  {Lon\ifmmode~\check{c}\else \v{c}\fi{}ar}, \citenamefont {Lukin},
  \citenamefont {Miller}, \citenamefont {Monroe}, \citenamefont {Nam},
  \citenamefont {Narang}, \citenamefont {Orcutt}, \citenamefont {Raymer},
  \citenamefont {Safavi-Naeini}, \citenamefont {Spiropulu}, \citenamefont
  {Srinivasan}, \citenamefont {Sun}, \citenamefont {Vu\ifmmode \check{c}\else
  \v{c}\fi{}kovi\ifmmode~\acute{c}\else \'{c}\fi{}}, \citenamefont {Waks},
  \citenamefont {Walsworth}, \citenamefont {Weiner},\ and\ \citenamefont
  {Zhang}}]{Awschalom2021}%
  \BibitemOpen
  \bibfield  {author} {\bibinfo {author} {\bibfnamefont {D.}~\bibnamefont
  {Awschalom}}, \bibinfo {author} {\bibfnamefont {K.~K.}\ \bibnamefont
  {Berggren}}, \bibinfo {author} {\bibfnamefont {H.}~\bibnamefont {Bernien}},
  \bibinfo {author} {\bibfnamefont {S.}~\bibnamefont {Bhave}}, \bibinfo
  {author} {\bibfnamefont {L.~D.}\ \bibnamefont {Carr}}, \bibinfo {author}
  {\bibfnamefont {P.}~\bibnamefont {Davids}}, \bibinfo {author} {\bibfnamefont
  {S.~E.}\ \bibnamefont {Economou}}, \bibinfo {author} {\bibfnamefont
  {D.}~\bibnamefont {Englund}}, \bibinfo {author} {\bibfnamefont
  {A.}~\bibnamefont {Faraon}}, \bibinfo {author} {\bibfnamefont
  {M.}~\bibnamefont {Fejer}}, \bibinfo {author} {\bibfnamefont
  {S.}~\bibnamefont {Guha}}, \bibinfo {author} {\bibfnamefont {M.~V.}\
  \bibnamefont {Gustafsson}}, \bibinfo {author} {\bibfnamefont
  {E.}~\bibnamefont {Hu}}, \bibinfo {author} {\bibfnamefont {L.}~\bibnamefont
  {Jiang}}, \bibinfo {author} {\bibfnamefont {J.}~\bibnamefont {Kim}}, \bibinfo
  {author} {\bibfnamefont {B.}~\bibnamefont {Korzh}}, \bibinfo {author}
  {\bibfnamefont {P.}~\bibnamefont {Kumar}}, \bibinfo {author} {\bibfnamefont
  {P.~G.}\ \bibnamefont {Kwiat}}, \bibinfo {author} {\bibfnamefont
  {M.}~\bibnamefont {Lon\ifmmode~\check{c}\else \v{c}\fi{}ar}}, \bibinfo
  {author} {\bibfnamefont {M.~D.}\ \bibnamefont {Lukin}}, \bibinfo {author}
  {\bibfnamefont {D.~A.}\ \bibnamefont {Miller}}, \bibinfo {author}
  {\bibfnamefont {C.}~\bibnamefont {Monroe}}, \bibinfo {author} {\bibfnamefont
  {S.~W.}\ \bibnamefont {Nam}}, \bibinfo {author} {\bibfnamefont
  {P.}~\bibnamefont {Narang}}, \bibinfo {author} {\bibfnamefont {J.~S.}\
  \bibnamefont {Orcutt}}, \bibinfo {author} {\bibfnamefont {M.~G.}\
  \bibnamefont {Raymer}}, \bibinfo {author} {\bibfnamefont {A.~H.}\
  \bibnamefont {Safavi-Naeini}}, \bibinfo {author} {\bibfnamefont
  {M.}~\bibnamefont {Spiropulu}}, \bibinfo {author} {\bibfnamefont
  {K.}~\bibnamefont {Srinivasan}}, \bibinfo {author} {\bibfnamefont
  {S.}~\bibnamefont {Sun}}, \bibinfo {author} {\bibfnamefont {J.}~\bibnamefont
  {Vu\ifmmode \check{c}\else \v{c}\fi{}kovi\ifmmode~\acute{c}\else
  \'{c}\fi{}}}, \bibinfo {author} {\bibfnamefont {E.}~\bibnamefont {Waks}},
  \bibinfo {author} {\bibfnamefont {R.}~\bibnamefont {Walsworth}}, \bibinfo
  {author} {\bibfnamefont {A.~M.}\ \bibnamefont {Weiner}},\ and\ \bibinfo
  {author} {\bibfnamefont {Z.}~\bibnamefont {Zhang}},\ }\bibfield  {title}
  {\bibinfo {title} {Development of quantum interconnects (quics) for
  next-generation information technologies},\ }\href
  {https://doi.org/10.1103/PRXQuantum.2.017002} {\bibfield  {journal} {\bibinfo
   {journal} {PRX Quantum}\ }\textbf {\bibinfo {volume} {2}},\ \bibinfo {pages}
  {017002} (\bibinfo {year} {2021})}\BibitemShut {NoStop}%
\bibitem [{\citenamefont {Bennett}\ \emph {et~al.}(1996)\citenamefont
  {Bennett}, \citenamefont {Brassard}, \citenamefont {Popescu}, \citenamefont
  {Schumacher}, \citenamefont {Smolin},\ and\ \citenamefont
  {Wootters}}]{bennett1996purification}%
  \BibitemOpen
  \bibfield  {author} {\bibinfo {author} {\bibfnamefont {C.~H.}\ \bibnamefont
  {Bennett}}, \bibinfo {author} {\bibfnamefont {G.}~\bibnamefont {Brassard}},
  \bibinfo {author} {\bibfnamefont {S.}~\bibnamefont {Popescu}}, \bibinfo
  {author} {\bibfnamefont {B.}~\bibnamefont {Schumacher}}, \bibinfo {author}
  {\bibfnamefont {J.~A.}\ \bibnamefont {Smolin}},\ and\ \bibinfo {author}
  {\bibfnamefont {W.~K.}\ \bibnamefont {Wootters}},\ }\bibfield  {title}
  {\bibinfo {title} {Purification of noisy entanglement and faithful
  teleportation via noisy channels},\ }\href@noop {} {\bibfield  {journal}
  {\bibinfo  {journal} {Physical review letters}\ }\textbf {\bibinfo {volume}
  {76}},\ \bibinfo {pages} {722} (\bibinfo {year} {1996})}\BibitemShut
  {NoStop}%
\bibitem [{\citenamefont {Deutsch}\ \emph {et~al.}(1996)\citenamefont
  {Deutsch}, \citenamefont {Ekert}, \citenamefont {Jozsa}, \citenamefont
  {Macchiavello}, \citenamefont {Popescu},\ and\ \citenamefont
  {Sanpera}}]{deutsch1996quantum}%
  \BibitemOpen
  \bibfield  {author} {\bibinfo {author} {\bibfnamefont {D.}~\bibnamefont
  {Deutsch}}, \bibinfo {author} {\bibfnamefont {A.}~\bibnamefont {Ekert}},
  \bibinfo {author} {\bibfnamefont {R.}~\bibnamefont {Jozsa}}, \bibinfo
  {author} {\bibfnamefont {C.}~\bibnamefont {Macchiavello}}, \bibinfo {author}
  {\bibfnamefont {S.}~\bibnamefont {Popescu}},\ and\ \bibinfo {author}
  {\bibfnamefont {A.}~\bibnamefont {Sanpera}},\ }\bibfield  {title} {\bibinfo
  {title} {Quantum privacy amplification and the security of quantum
  cryptography over noisy channels},\ }\href@noop {} {\bibfield  {journal}
  {\bibinfo  {journal} {Physical review letters}\ }\textbf {\bibinfo {volume}
  {77}},\ \bibinfo {pages} {2818} (\bibinfo {year} {1996})}\BibitemShut
  {NoStop}%
\bibitem [{\citenamefont {Fowler}\ \emph {et~al.}(2010)\citenamefont {Fowler},
  \citenamefont {Wang}, \citenamefont {Hill}, \citenamefont {Ladd},
  \citenamefont {Van~Meter},\ and\ \citenamefont
  {Hollenberg}}]{fowler2010surface}%
  \BibitemOpen
  \bibfield  {author} {\bibinfo {author} {\bibfnamefont {A.~G.}\ \bibnamefont
  {Fowler}}, \bibinfo {author} {\bibfnamefont {D.~S.}\ \bibnamefont {Wang}},
  \bibinfo {author} {\bibfnamefont {C.~D.}\ \bibnamefont {Hill}}, \bibinfo
  {author} {\bibfnamefont {T.~D.}\ \bibnamefont {Ladd}}, \bibinfo {author}
  {\bibfnamefont {R.}~\bibnamefont {Van~Meter}},\ and\ \bibinfo {author}
  {\bibfnamefont {L.~C.}\ \bibnamefont {Hollenberg}},\ }\bibfield  {title}
  {\bibinfo {title} {Surface code quantum communication},\ }\href@noop {}
  {\bibfield  {journal} {\bibinfo  {journal} {Physical review letters}\
  }\textbf {\bibinfo {volume} {104}},\ \bibinfo {pages} {180503} (\bibinfo
  {year} {2010})}\BibitemShut {NoStop}%
\bibitem [{\citenamefont {Javadi-Abhari}\ \emph {et~al.}(2017)\citenamefont
  {Javadi-Abhari}, \citenamefont {Gokhale}, \citenamefont {Holmes},
  \citenamefont {Franklin}, \citenamefont {Brown}, \citenamefont {Martonosi},\
  and\ \citenamefont {Chong}}]{javadi2017optimized}%
  \BibitemOpen
  \bibfield  {author} {\bibinfo {author} {\bibfnamefont {A.}~\bibnamefont
  {Javadi-Abhari}}, \bibinfo {author} {\bibfnamefont {P.}~\bibnamefont
  {Gokhale}}, \bibinfo {author} {\bibfnamefont {A.}~\bibnamefont {Holmes}},
  \bibinfo {author} {\bibfnamefont {D.}~\bibnamefont {Franklin}}, \bibinfo
  {author} {\bibfnamefont {K.~R.}\ \bibnamefont {Brown}}, \bibinfo {author}
  {\bibfnamefont {M.}~\bibnamefont {Martonosi}},\ and\ \bibinfo {author}
  {\bibfnamefont {F.~T.}\ \bibnamefont {Chong}},\ }\bibfield  {title} {\bibinfo
  {title} {Optimized surface code communication in superconducting quantum
  computers},\ }in\ \href@noop {} {\emph {\bibinfo {booktitle} {Proceedings of
  the 50th Annual IEEE/ACM International Symposium on Microarchitecture}}}\
  (\bibinfo {year} {2017})\ pp.\ \bibinfo {pages} {692--705}\BibitemShut
  {NoStop}%
\bibitem [{\citenamefont {Zheng}\ \emph {et~al.}(2022)\citenamefont {Zheng},
  \citenamefont {Sharma},\ and\ \citenamefont
  {Borregaard}}]{zheng2022entanglement}%
  \BibitemOpen
  \bibfield  {author} {\bibinfo {author} {\bibfnamefont {Y.}~\bibnamefont
  {Zheng}}, \bibinfo {author} {\bibfnamefont {H.}~\bibnamefont {Sharma}},\ and\
  \bibinfo {author} {\bibfnamefont {J.}~\bibnamefont {Borregaard}},\ }\bibfield
   {title} {\bibinfo {title} {Entanglement distribution with minimal memory
  requirements using time-bin photonic qudits},\ }\href@noop {} {\bibfield
  {journal} {\bibinfo  {journal} {PRX Quantum}\ }\textbf {\bibinfo {volume}
  {3}},\ \bibinfo {pages} {040319} (\bibinfo {year} {2022})}\BibitemShut
  {NoStop}%
\bibitem [{\citenamefont {Xie}\ \emph {et~al.}(2021)\citenamefont {Xie},
  \citenamefont {Liu}, \citenamefont {Mo}, \citenamefont {Li},\ and\
  \citenamefont {Li}}]{xie2021quantum}%
  \BibitemOpen
  \bibfield  {author} {\bibinfo {author} {\bibfnamefont {Z.}~\bibnamefont
  {Xie}}, \bibinfo {author} {\bibfnamefont {Y.}~\bibnamefont {Liu}}, \bibinfo
  {author} {\bibfnamefont {X.}~\bibnamefont {Mo}}, \bibinfo {author}
  {\bibfnamefont {T.}~\bibnamefont {Li}},\ and\ \bibinfo {author}
  {\bibfnamefont {Z.}~\bibnamefont {Li}},\ }\bibfield  {title} {\bibinfo
  {title} {Quantum entanglement creation for distant quantum memories via
  time-bin multiplexing},\ }\href@noop {} {\bibfield  {journal} {\bibinfo
  {journal} {Physical Review A}\ }\textbf {\bibinfo {volume} {104}},\ \bibinfo
  {pages} {062409} (\bibinfo {year} {2021})}\BibitemShut {NoStop}%
\bibitem [{\citenamefont {Piparo}\ \emph {et~al.}(2019)\citenamefont {Piparo},
  \citenamefont {Munro},\ and\ \citenamefont {Nemoto}}]{piparo2019quantum}%
  \BibitemOpen
  \bibfield  {author} {\bibinfo {author} {\bibfnamefont {N.~L.}\ \bibnamefont
  {Piparo}}, \bibinfo {author} {\bibfnamefont {W.~J.}\ \bibnamefont {Munro}},\
  and\ \bibinfo {author} {\bibfnamefont {K.}~\bibnamefont {Nemoto}},\
  }\bibfield  {title} {\bibinfo {title} {Quantum multiplexing},\ }\href@noop {}
  {\bibfield  {journal} {\bibinfo  {journal} {Physical Review A}\ }\textbf
  {\bibinfo {volume} {99}},\ \bibinfo {pages} {022337} (\bibinfo {year}
  {2019})}\BibitemShut {NoStop}%
\bibitem [{\citenamefont {Zhou}\ \emph {et~al.}(2023)\citenamefont {Zhou},
  \citenamefont {Li}, \citenamefont {Xia} \emph {et~al.}}]{zhou2023parallel}%
  \BibitemOpen
  \bibfield  {author} {\bibinfo {author} {\bibfnamefont {H.}~\bibnamefont
  {Zhou}}, \bibinfo {author} {\bibfnamefont {T.}~\bibnamefont {Li}}, \bibinfo
  {author} {\bibfnamefont {K.}~\bibnamefont {Xia}}, \emph {et~al.},\ }\bibfield
   {title} {\bibinfo {title} {Parallel and heralded multiqubit entanglement
  generation for quantum networks},\ }\href@noop {} {\bibfield  {journal}
  {\bibinfo  {journal} {Physical Review A}\ }\textbf {\bibinfo {volume}
  {107}},\ \bibinfo {pages} {022428} (\bibinfo {year} {2023})}\BibitemShut
  {NoStop}%
\bibitem [{\citenamefont {Krastanov}\ \emph {et~al.}(2019)\citenamefont
  {Krastanov}, \citenamefont {Albert},\ and\ \citenamefont
  {Jiang}}]{krastanov2019optimized}%
  \BibitemOpen
  \bibfield  {author} {\bibinfo {author} {\bibfnamefont {S.}~\bibnamefont
  {Krastanov}}, \bibinfo {author} {\bibfnamefont {V.~V.}\ \bibnamefont
  {Albert}},\ and\ \bibinfo {author} {\bibfnamefont {L.}~\bibnamefont
  {Jiang}},\ }\bibfield  {title} {\bibinfo {title} {Optimized entanglement
  purification},\ }\href@noop {} {\bibfield  {journal} {\bibinfo  {journal}
  {Quantum}\ }\textbf {\bibinfo {volume} {3}},\ \bibinfo {pages} {123}
  (\bibinfo {year} {2019})}\BibitemShut {NoStop}%
\bibitem [{\citenamefont {Knill}\ \emph {et~al.}(2001)\citenamefont {Knill},
  \citenamefont {Laflamme},\ and\ \citenamefont {Milburn}}]{knill2001scheme}%
  \BibitemOpen
  \bibfield  {author} {\bibinfo {author} {\bibfnamefont {E.}~\bibnamefont
  {Knill}}, \bibinfo {author} {\bibfnamefont {R.}~\bibnamefont {Laflamme}},\
  and\ \bibinfo {author} {\bibfnamefont {G.~J.}\ \bibnamefont {Milburn}},\
  }\bibfield  {title} {\bibinfo {title} {A scheme for efficient quantum
  computation with linear optics},\ }\href@noop {} {\bibfield  {journal}
  {\bibinfo  {journal} {nature}\ }\textbf {\bibinfo {volume} {409}},\ \bibinfo
  {pages} {46} (\bibinfo {year} {2001})}\BibitemShut {NoStop}%
\bibitem [{\citenamefont {Knill}(2005)}]{knill2005quantum}%
  \BibitemOpen
  \bibfield  {author} {\bibinfo {author} {\bibfnamefont {E.}~\bibnamefont
  {Knill}},\ }\bibfield  {title} {\bibinfo {title} {Quantum computing with
  realistically noisy devices},\ }\href@noop {} {\bibfield  {journal} {\bibinfo
   {journal} {Nature}\ }\textbf {\bibinfo {volume} {434}},\ \bibinfo {pages}
  {39} (\bibinfo {year} {2005})}\BibitemShut {NoStop}%
\bibitem [{\citenamefont {Namiki}\ \emph {et~al.}(2016)\citenamefont {Namiki},
  \citenamefont {Jiang}, \citenamefont {Kim},\ and\ \citenamefont
  {L{\"u}tkenhaus}}]{namiki2016role}%
  \BibitemOpen
  \bibfield  {author} {\bibinfo {author} {\bibfnamefont {R.}~\bibnamefont
  {Namiki}}, \bibinfo {author} {\bibfnamefont {L.}~\bibnamefont {Jiang}},
  \bibinfo {author} {\bibfnamefont {J.}~\bibnamefont {Kim}},\ and\ \bibinfo
  {author} {\bibfnamefont {N.}~\bibnamefont {L{\"u}tkenhaus}},\ }\bibfield
  {title} {\bibinfo {title} {Role of syndrome information on a one-way quantum
  repeater using teleportation-based error correction},\ }\href@noop {}
  {\bibfield  {journal} {\bibinfo  {journal} {Physical Review A}\ }\textbf
  {\bibinfo {volume} {94}},\ \bibinfo {pages} {052304} (\bibinfo {year}
  {2016})}\BibitemShut {NoStop}%
\bibitem [{\citenamefont {Devitt}\ \emph {et~al.}(2013)\citenamefont {Devitt},
  \citenamefont {Munro},\ and\ \citenamefont {Nemoto}}]{devitt2013quantum}%
  \BibitemOpen
  \bibfield  {author} {\bibinfo {author} {\bibfnamefont {S.~J.}\ \bibnamefont
  {Devitt}}, \bibinfo {author} {\bibfnamefont {W.~J.}\ \bibnamefont {Munro}},\
  and\ \bibinfo {author} {\bibfnamefont {K.}~\bibnamefont {Nemoto}},\
  }\bibfield  {title} {\bibinfo {title} {Quantum error correction for
  beginners},\ }\href@noop {} {\bibfield  {journal} {\bibinfo  {journal}
  {Reports on Progress in Physics}\ }\textbf {\bibinfo {volume} {76}},\
  \bibinfo {pages} {076001} (\bibinfo {year} {2013})}\BibitemShut {NoStop}%
\bibitem [{\citenamefont {Knall}\ \emph
  {et~al.}(2022{\natexlab{a}})\citenamefont {Knall}, \citenamefont {Knaut},
  \citenamefont {Bekenstein}, \citenamefont {Assumpcao}, \citenamefont
  {Stroganov}, \citenamefont {Gong}, \citenamefont {Huan}, \citenamefont
  {Stas}, \citenamefont {Machielse}, \citenamefont {Chalupnik}, \citenamefont
  {Levonian}, \citenamefont {Suleymanzade}, \citenamefont {Riedinger},
  \citenamefont {Park}, \citenamefont {Lon\ifmmode~\check{c}\else
  \v{c}\fi{}ar}, \citenamefont {Bhaskar},\ and\ \citenamefont
  {Lukin}}]{Knall2022}%
  \BibitemOpen
  \bibfield  {author} {\bibinfo {author} {\bibfnamefont {E.~N.}\ \bibnamefont
  {Knall}}, \bibinfo {author} {\bibfnamefont {C.~M.}\ \bibnamefont {Knaut}},
  \bibinfo {author} {\bibfnamefont {R.}~\bibnamefont {Bekenstein}}, \bibinfo
  {author} {\bibfnamefont {D.~R.}\ \bibnamefont {Assumpcao}}, \bibinfo {author}
  {\bibfnamefont {P.~L.}\ \bibnamefont {Stroganov}}, \bibinfo {author}
  {\bibfnamefont {W.}~\bibnamefont {Gong}}, \bibinfo {author} {\bibfnamefont
  {Y.~Q.}\ \bibnamefont {Huan}}, \bibinfo {author} {\bibfnamefont {P.-J.}\
  \bibnamefont {Stas}}, \bibinfo {author} {\bibfnamefont {B.}~\bibnamefont
  {Machielse}}, \bibinfo {author} {\bibfnamefont {M.}~\bibnamefont
  {Chalupnik}}, \bibinfo {author} {\bibfnamefont {D.}~\bibnamefont {Levonian}},
  \bibinfo {author} {\bibfnamefont {A.}~\bibnamefont {Suleymanzade}}, \bibinfo
  {author} {\bibfnamefont {R.}~\bibnamefont {Riedinger}}, \bibinfo {author}
  {\bibfnamefont {H.}~\bibnamefont {Park}}, \bibinfo {author} {\bibfnamefont
  {M.}~\bibnamefont {Lon\ifmmode~\check{c}\else \v{c}\fi{}ar}}, \bibinfo
  {author} {\bibfnamefont {M.~K.}\ \bibnamefont {Bhaskar}},\ and\ \bibinfo
  {author} {\bibfnamefont {M.~D.}\ \bibnamefont {Lukin}},\ }\bibfield  {title}
  {\bibinfo {title} {Efficient source of shaped single photons based on an
  integrated diamond nanophotonic system},\ }\href
  {https://doi.org/10.1103/PhysRevLett.129.053603} {\bibfield  {journal}
  {\bibinfo  {journal} {Phys. Rev. Lett.}\ }\textbf {\bibinfo {volume} {129}},\
  \bibinfo {pages} {053603} (\bibinfo {year} {2022}{\natexlab{a}})}\BibitemShut
  {NoStop}%
\bibitem [{\citenamefont {Duan}\ and\ \citenamefont
  {Kimble}(2004)}]{duan2004scalable}%
  \BibitemOpen
  \bibfield  {author} {\bibinfo {author} {\bibfnamefont {L.-M.}\ \bibnamefont
  {Duan}}\ and\ \bibinfo {author} {\bibfnamefont {H.}~\bibnamefont {Kimble}},\
  }\bibfield  {title} {\bibinfo {title} {Scalable photonic quantum computation
  through cavity-assisted interactions},\ }\href@noop {} {\bibfield  {journal}
  {\bibinfo  {journal} {Physical review letters}\ }\textbf {\bibinfo {volume}
  {92}},\ \bibinfo {pages} {127902} (\bibinfo {year} {2004})}\BibitemShut
  {NoStop}%
\bibitem [{\citenamefont {Bhaskar}\ \emph {et~al.}(2020)\citenamefont
  {Bhaskar}, \citenamefont {Riedinger}, \citenamefont {Machielse},
  \citenamefont {Levonian}, \citenamefont {Nguyen}, \citenamefont {Knall},
  \citenamefont {Park}, \citenamefont {Englund}, \citenamefont {Lon{\v{c}}ar},
  \citenamefont {Sukachev} \emph {et~al.}}]{bhaskar2020experimental}%
  \BibitemOpen
  \bibfield  {author} {\bibinfo {author} {\bibfnamefont {M.~K.}\ \bibnamefont
  {Bhaskar}}, \bibinfo {author} {\bibfnamefont {R.}~\bibnamefont {Riedinger}},
  \bibinfo {author} {\bibfnamefont {B.}~\bibnamefont {Machielse}}, \bibinfo
  {author} {\bibfnamefont {D.~S.}\ \bibnamefont {Levonian}}, \bibinfo {author}
  {\bibfnamefont {C.~T.}\ \bibnamefont {Nguyen}}, \bibinfo {author}
  {\bibfnamefont {E.~N.}\ \bibnamefont {Knall}}, \bibinfo {author}
  {\bibfnamefont {H.}~\bibnamefont {Park}}, \bibinfo {author} {\bibfnamefont
  {D.}~\bibnamefont {Englund}}, \bibinfo {author} {\bibfnamefont
  {M.}~\bibnamefont {Lon{\v{c}}ar}}, \bibinfo {author} {\bibfnamefont {D.~D.}\
  \bibnamefont {Sukachev}}, \emph {et~al.},\ }\bibfield  {title} {\bibinfo
  {title} {Experimental demonstration of memory-enhanced quantum
  communication},\ }\href@noop {} {\bibfield  {journal} {\bibinfo  {journal}
  {Nature}\ }\textbf {\bibinfo {volume} {580}},\ \bibinfo {pages} {60}
  (\bibinfo {year} {2020})}\BibitemShut {NoStop}%
\bibitem [{\citenamefont {D{\"u}r}\ and\ \citenamefont
  {Briegel}(2007)}]{dur2007entanglement}%
  \BibitemOpen
  \bibfield  {author} {\bibinfo {author} {\bibfnamefont {W.}~\bibnamefont
  {D{\"u}r}}\ and\ \bibinfo {author} {\bibfnamefont {H.~J.}\ \bibnamefont
  {Briegel}},\ }\bibfield  {title} {\bibinfo {title} {Entanglement purification
  and quantum error correction},\ }\href@noop {} {\bibfield  {journal}
  {\bibinfo  {journal} {Reports on Progress in Physics}\ }\textbf {\bibinfo
  {volume} {70}},\ \bibinfo {pages} {1381} (\bibinfo {year}
  {2007})}\BibitemShut {NoStop}%
\bibitem [{\citenamefont {Muralidharan}\ \emph {et~al.}(2016)\citenamefont
  {Muralidharan}, \citenamefont {Li}, \citenamefont {Kim}, \citenamefont
  {L{\"u}tkenhaus}, \citenamefont {Lukin},\ and\ \citenamefont
  {Jiang}}]{muralidharan2016optimal}%
  \BibitemOpen
  \bibfield  {author} {\bibinfo {author} {\bibfnamefont {S.}~\bibnamefont
  {Muralidharan}}, \bibinfo {author} {\bibfnamefont {L.}~\bibnamefont {Li}},
  \bibinfo {author} {\bibfnamefont {J.}~\bibnamefont {Kim}}, \bibinfo {author}
  {\bibfnamefont {N.}~\bibnamefont {L{\"u}tkenhaus}}, \bibinfo {author}
  {\bibfnamefont {M.~D.}\ \bibnamefont {Lukin}},\ and\ \bibinfo {author}
  {\bibfnamefont {L.}~\bibnamefont {Jiang}},\ }\bibfield  {title} {\bibinfo
  {title} {Optimal architectures for long distance quantum communication},\
  }\href@noop {} {\bibfield  {journal} {\bibinfo  {journal} {Scientific
  reports}\ }\textbf {\bibinfo {volume} {6}},\ \bibinfo {pages} {20463}
  (\bibinfo {year} {2016})}\BibitemShut {NoStop}%
\bibitem [{\citenamefont {Monroe}\ \emph {et~al.}(2014)\citenamefont {Monroe},
  \citenamefont {Raussendorf}, \citenamefont {Ruthven}, \citenamefont {Brown},
  \citenamefont {Maunz}, \citenamefont {Duan},\ and\ \citenamefont
  {Kim}}]{monroe2014large}%
  \BibitemOpen
  \bibfield  {author} {\bibinfo {author} {\bibfnamefont {C.}~\bibnamefont
  {Monroe}}, \bibinfo {author} {\bibfnamefont {R.}~\bibnamefont {Raussendorf}},
  \bibinfo {author} {\bibfnamefont {A.}~\bibnamefont {Ruthven}}, \bibinfo
  {author} {\bibfnamefont {K.~R.}\ \bibnamefont {Brown}}, \bibinfo {author}
  {\bibfnamefont {P.}~\bibnamefont {Maunz}}, \bibinfo {author} {\bibfnamefont
  {L.-M.}\ \bibnamefont {Duan}},\ and\ \bibinfo {author} {\bibfnamefont
  {J.}~\bibnamefont {Kim}},\ }\bibfield  {title} {\bibinfo {title} {Large-scale
  modular quantum-computer architecture with atomic memory and photonic
  interconnects},\ }\href@noop {} {\bibfield  {journal} {\bibinfo  {journal}
  {Physical Review A}\ }\textbf {\bibinfo {volume} {89}},\ \bibinfo {pages}
  {022317} (\bibinfo {year} {2014})}\BibitemShut {NoStop}%
\bibitem [{\citenamefont {Zhang}\ and\ \citenamefont
  {Man}(2005)}]{zhang2005many}%
  \BibitemOpen
  \bibfield  {author} {\bibinfo {author} {\bibfnamefont {Z.-J.}\ \bibnamefont
  {Zhang}}\ and\ \bibinfo {author} {\bibfnamefont {Z.-X.}\ \bibnamefont
  {Man}},\ }\bibfield  {title} {\bibinfo {title} {Many-agent controlled
  teleportation of multi-qubit quantum information},\ }\href@noop {} {\bibfield
   {journal} {\bibinfo  {journal} {Physics Letters A}\ }\textbf {\bibinfo
  {volume} {341}},\ \bibinfo {pages} {55} (\bibinfo {year} {2005})}\BibitemShut
  {NoStop}%
\bibitem [{\citenamefont {Li}\ \emph {et~al.}(2016)\citenamefont {Li},
  \citenamefont {Li}, \citenamefont {Nie},\ and\ \citenamefont
  {Sang}}]{li2016quantum}%
  \BibitemOpen
  \bibfield  {author} {\bibinfo {author} {\bibfnamefont {Y.-h.}\ \bibnamefont
  {Li}}, \bibinfo {author} {\bibfnamefont {X.-l.}\ \bibnamefont {Li}}, \bibinfo
  {author} {\bibfnamefont {L.-p.}\ \bibnamefont {Nie}},\ and\ \bibinfo {author}
  {\bibfnamefont {M.-h.}\ \bibnamefont {Sang}},\ }\bibfield  {title} {\bibinfo
  {title} {Quantum teleportation of three and four-qubit state using
  multi-qubit cluster states},\ }\href@noop {} {\bibfield  {journal} {\bibinfo
  {journal} {International Journal of Theoretical Physics}\ }\textbf {\bibinfo
  {volume} {55}},\ \bibinfo {pages} {1820} (\bibinfo {year}
  {2016})}\BibitemShut {NoStop}%
\bibitem [{\citenamefont {Laflamme}\ \emph {et~al.}(1996)\citenamefont
  {Laflamme}, \citenamefont {Miquel}, \citenamefont {Paz},\ and\ \citenamefont
  {Zurek}}]{laflamme1996perfect}%
  \BibitemOpen
  \bibfield  {author} {\bibinfo {author} {\bibfnamefont {R.}~\bibnamefont
  {Laflamme}}, \bibinfo {author} {\bibfnamefont {C.}~\bibnamefont {Miquel}},
  \bibinfo {author} {\bibfnamefont {J.~P.}\ \bibnamefont {Paz}},\ and\ \bibinfo
  {author} {\bibfnamefont {W.~H.}\ \bibnamefont {Zurek}},\ }\bibfield  {title}
  {\bibinfo {title} {Perfect quantum error correcting code},\ }\href@noop {}
  {\bibfield  {journal} {\bibinfo  {journal} {Physical Review Letters}\
  }\textbf {\bibinfo {volume} {77}},\ \bibinfo {pages} {198} (\bibinfo {year}
  {1996})}\BibitemShut {NoStop}%
\bibitem [{\citenamefont {Grassl}\ \emph {et~al.}(1997)\citenamefont {Grassl},
  \citenamefont {Beth},\ and\ \citenamefont {Pellizzari}}]{grassl1997codes}%
  \BibitemOpen
  \bibfield  {author} {\bibinfo {author} {\bibfnamefont {M.}~\bibnamefont
  {Grassl}}, \bibinfo {author} {\bibfnamefont {T.}~\bibnamefont {Beth}},\ and\
  \bibinfo {author} {\bibfnamefont {T.}~\bibnamefont {Pellizzari}},\ }\bibfield
   {title} {\bibinfo {title} {Codes for the quantum erasure channel},\
  }\href@noop {} {\bibfield  {journal} {\bibinfo  {journal} {Physical Review
  A}\ }\textbf {\bibinfo {volume} {56}},\ \bibinfo {pages} {33} (\bibinfo
  {year} {1997})}\BibitemShut {NoStop}%
\bibitem [{\citenamefont {Lee}\ \emph {et~al.}(2018)\citenamefont {Lee},
  \citenamefont {Wells}, \citenamefont {Villa}, \citenamefont {Kalliakos},
  \citenamefont {Stevenson}, \citenamefont {Ellis}, \citenamefont {Farrer},
  \citenamefont {Ritchie}, \citenamefont {Bennett},\ and\ \citenamefont
  {Shields}}]{lee2018controllable}%
  \BibitemOpen
  \bibfield  {author} {\bibinfo {author} {\bibfnamefont {J.}~\bibnamefont
  {Lee}}, \bibinfo {author} {\bibfnamefont {L.}~\bibnamefont {Wells}}, \bibinfo
  {author} {\bibfnamefont {B.}~\bibnamefont {Villa}}, \bibinfo {author}
  {\bibfnamefont {S.}~\bibnamefont {Kalliakos}}, \bibinfo {author}
  {\bibfnamefont {R.}~\bibnamefont {Stevenson}}, \bibinfo {author}
  {\bibfnamefont {D.}~\bibnamefont {Ellis}}, \bibinfo {author} {\bibfnamefont
  {I.}~\bibnamefont {Farrer}}, \bibinfo {author} {\bibfnamefont
  {D.}~\bibnamefont {Ritchie}}, \bibinfo {author} {\bibfnamefont
  {A.}~\bibnamefont {Bennett}},\ and\ \bibinfo {author} {\bibfnamefont
  {A.}~\bibnamefont {Shields}},\ }\bibfield  {title} {\bibinfo {title}
  {Controllable photonic time-bin qubits from a quantum dot},\ }\href@noop {}
  {\bibfield  {journal} {\bibinfo  {journal} {Physical Review X}\ }\textbf
  {\bibinfo {volume} {8}},\ \bibinfo {pages} {021078} (\bibinfo {year}
  {2018})}\BibitemShut {NoStop}%
\bibitem [{\citenamefont {Knall}\ \emph
  {et~al.}(2022{\natexlab{b}})\citenamefont {Knall}, \citenamefont {Knaut},
  \citenamefont {Bekenstein}, \citenamefont {Assumpcao}, \citenamefont
  {Stroganov}, \citenamefont {Gong}, \citenamefont {Huan}, \citenamefont
  {Stas}, \citenamefont {Machielse}, \citenamefont {Chalupnik} \emph
  {et~al.}}]{knall2022efficient}%
  \BibitemOpen
  \bibfield  {author} {\bibinfo {author} {\bibfnamefont {E.~N.}\ \bibnamefont
  {Knall}}, \bibinfo {author} {\bibfnamefont {C.~M.}\ \bibnamefont {Knaut}},
  \bibinfo {author} {\bibfnamefont {R.}~\bibnamefont {Bekenstein}}, \bibinfo
  {author} {\bibfnamefont {D.~R.}\ \bibnamefont {Assumpcao}}, \bibinfo {author}
  {\bibfnamefont {P.~L.}\ \bibnamefont {Stroganov}}, \bibinfo {author}
  {\bibfnamefont {W.}~\bibnamefont {Gong}}, \bibinfo {author} {\bibfnamefont
  {Y.~Q.}\ \bibnamefont {Huan}}, \bibinfo {author} {\bibfnamefont {P.-J.}\
  \bibnamefont {Stas}}, \bibinfo {author} {\bibfnamefont {B.}~\bibnamefont
  {Machielse}}, \bibinfo {author} {\bibfnamefont {M.}~\bibnamefont
  {Chalupnik}}, \emph {et~al.},\ }\bibfield  {title} {\bibinfo {title}
  {Efficient source of shaped single photons based on an integrated diamond
  nanophotonic system},\ }\href@noop {} {\bibfield  {journal} {\bibinfo
  {journal} {Physical Review Letters}\ }\textbf {\bibinfo {volume} {129}},\
  \bibinfo {pages} {053603} (\bibinfo {year} {2022}{\natexlab{b}})}\BibitemShut
  {NoStop}%
\bibitem [{\citenamefont {Sharma}()}]{hemant}%
  \BibitemOpen
  \bibfield  {author} {\bibinfo {author} {\bibfnamefont {H.}~\bibnamefont
  {Sharma}},\ }\href@noop {} {\bibinfo {title} {Scattering based scheme for
  generation of photonic tree-cluster states}},\ \bibinfo {howpublished}
  {http://resolver.tudelft.nl/uuid:64e71ea0-2f72-4f35-b253-8a80326fd7af}\BibitemShut
  {NoStop}%
\bibitem [{\citenamefont {Tiurev}\ \emph {et~al.}(2021)\citenamefont {Tiurev},
  \citenamefont {Mirambell}, \citenamefont {Lauritzen}, \citenamefont {Appel},
  \citenamefont {Tiranov}, \citenamefont {Lodahl},\ and\ \citenamefont
  {S{\o}rensen}}]{tiurev2021fidelity}%
  \BibitemOpen
  \bibfield  {author} {\bibinfo {author} {\bibfnamefont {K.}~\bibnamefont
  {Tiurev}}, \bibinfo {author} {\bibfnamefont {P.~L.}\ \bibnamefont
  {Mirambell}}, \bibinfo {author} {\bibfnamefont {M.~B.}\ \bibnamefont
  {Lauritzen}}, \bibinfo {author} {\bibfnamefont {M.~H.}\ \bibnamefont
  {Appel}}, \bibinfo {author} {\bibfnamefont {A.}~\bibnamefont {Tiranov}},
  \bibinfo {author} {\bibfnamefont {P.}~\bibnamefont {Lodahl}},\ and\ \bibinfo
  {author} {\bibfnamefont {A.~S.}\ \bibnamefont {S{\o}rensen}},\ }\bibfield
  {title} {\bibinfo {title} {Fidelity of time-bin-entangled multiphoton states
  from a quantum emitter},\ }\href@noop {} {\bibfield  {journal} {\bibinfo
  {journal} {Physical Review A}\ }\textbf {\bibinfo {volume} {104}},\ \bibinfo
  {pages} {052604} (\bibinfo {year} {2021})}\BibitemShut {NoStop}%
\bibitem [{\citenamefont {Yang}\ \emph {et~al.}(2004)\citenamefont {Yang},
  \citenamefont {Chu},\ and\ \citenamefont {Han}}]{yang2004simplified}%
  \BibitemOpen
  \bibfield  {author} {\bibinfo {author} {\bibfnamefont {C.-P.}\ \bibnamefont
  {Yang}}, \bibinfo {author} {\bibfnamefont {S.-I.}\ \bibnamefont {Chu}},\ and\
  \bibinfo {author} {\bibfnamefont {S.}~\bibnamefont {Han}},\ }\bibfield
  {title} {\bibinfo {title} {Simplified realization of two-qubit quantum phase
  gate with four-level systems in cavity qed},\ }\href@noop {} {\bibfield
  {journal} {\bibinfo  {journal} {Physical Review A}\ }\textbf {\bibinfo
  {volume} {70}},\ \bibinfo {pages} {044303} (\bibinfo {year}
  {2004})}\BibitemShut {NoStop}%
\bibitem [{dbl()}]{dbloss}%
  \BibitemOpen
  \href@noop {} {\bibinfo {title} {Thorlabs fiber parameters}},\ \bibinfo
  {howpublished} {https://www.thorlabs.com}\BibitemShut {NoStop}%
\bibitem [{\citenamefont {Li}\ and\ \citenamefont
  {Hayashi}(2020)}]{li2020advances}%
  \BibitemOpen
  \bibfield  {author} {\bibinfo {author} {\bibfnamefont {M.-J.}\ \bibnamefont
  {Li}}\ and\ \bibinfo {author} {\bibfnamefont {T.}~\bibnamefont {Hayashi}},\
  }\bibfield  {title} {\bibinfo {title} {Advances in low-loss, large-area, and
  multicore fibers},\ }in\ \href@noop {} {\emph {\bibinfo {booktitle} {Optical
  Fiber Telecommunications VII}}}\ (\bibinfo  {publisher} {Elsevier},\ \bibinfo
  {year} {2020})\ pp.\ \bibinfo {pages} {3--50}\BibitemShut {NoStop}%
\bibitem [{\citenamefont {Geller}\ and\ \citenamefont
  {Zhou}(2013)}]{geller2013efficient}%
  \BibitemOpen
  \bibfield  {author} {\bibinfo {author} {\bibfnamefont {M.~R.}\ \bibnamefont
  {Geller}}\ and\ \bibinfo {author} {\bibfnamefont {Z.}~\bibnamefont {Zhou}},\
  }\bibfield  {title} {\bibinfo {title} {Efficient error models for
  fault-tolerant architectures and the pauli twirling approximation},\
  }\href@noop {} {\bibfield  {journal} {\bibinfo  {journal} {Physical Review
  A}\ }\textbf {\bibinfo {volume} {88}},\ \bibinfo {pages} {012314} (\bibinfo
  {year} {2013})}\BibitemShut {NoStop}%
\end{thebibliography}%

\end{document}